# EXPLORING THE POTENTIAL OF CONVERSATIONAL AI SUPPORT FOR AGENT-BASED SOCIAL SIMULATION MODEL DESIGN

Peer-Olaf Siebers

School of Computer Science, University of Nottingham, Nottingham, UK
peer-olaf.siebers@nottingham.ac.uk

## ABSTRACT

ChatGPT, the AI-powered chatbot with a massive user base of hundreds of millions, has become a global phenomenon. However, the use of Conversational AI Systems (CAISs) like ChatGPT for research in the field of Social Simulation is still limited. Specifically, there is no evidence of its usage in Agent-Based Social Simulation (ABSS) model design. This paper takes a crucial first step toward exploring the untapped potential of this emerging technology in the context of ABSS model design. The research presented here demonstrates how CAISs can facilitate the development of innovative conceptual ABSS models in a concise timeframe and with minimal required upfront case-based knowledge. By employing advanced prompt engineering techniques and adhering to the Engineering ABSS framework, we have constructed a comprehensive prompt script that enables the design of conceptual ABSS models with or by the CAIS. A proof-of-concept application of the prompt script, used to generate the conceptual ABSS model for a case study on the impact of adaptive architecture in a museum environment, illustrates the practicality of the approach. Despite occasional inaccuracies and conversational divergence, the CAIS proved to be a valuable companion for ABSS modellers.

## ABBREVIATIONS USED

- ABM Agent-Based Model(ling)
- ABSS Agent-Based Social Simulation
- EABSS Engineering Agent-Based Social Simulation
- CAI Conversational AI
- CAIS Conversational AI System
- GABM Generative Agent-Based Model(ling)
- LLM Large Language Model
- NLG Natural Language Generation
- NLP Natural Language Processing

## 1 MOTIVATION

Developing innovative conceptual Agent-Based Social Simulation (ABSS) models with the involvement of relevant key stakeholders can be challenging. In response to this challenge, we created the Engineering Agent-Based Social Simulation (EABSS) framework some years ago, which supports the process of co-creating conceptual ABSS models in a structured and standardised manner, involving all relevant key stakeholders (Siebers and Klügl 2017). Despite our belief in the value of this resource for the social simulation community, its adoption has not met our expectations. Feedback gathered through discussions at conferences with community members pointed to several usability challenges that have limited broader engagement with the framework. These include a steep learning curve associated with effectively using the framework, the demanding role of the focus group moderator, and difficulties in gathering all necessary stakeholders for the process.

To address these concerns, our ongoing research efforts are directed at improving the usability of the EABSS framework, which we approach from two different angles. We are working on an Iteration and Information Re-Use Schema that guides the EABSS framework user on how to reuse information gathered in the previous steps when initiating the next step. This should ease the pressure put on the moderator when moving to the next step defined in the framework. Additionally, we are exploring the option of providing support to users through the use of chatbots that possess the ability to comprehend and generate text resembling human language. With the recent advancement of Conversational AI Systems (CAISs), there are numerous opportunities to assist the EABSS framework user throughout the conceptual model development process.

In this paper we focus on two of these opportunities: (1) providing the user with some domain knowledge and conceptual modelling ideas by conducting a dry run through the EABSS framework with the help of a CAIS, and (2) using roleplay games with virtual stakeholders that possess specific domain knowledge, to imitate the discussions taking place between real stakeholders during a co-creation session. Besides, we are exploring the role of CAISs as a brainstorming buddy and innovative idea generator in cases where there is no clear idea on how to get started with designing a model for a planned study.
The research presented here was conducted between January and May 2024, using the CAISs available at that time. Our work introduces a novel approach for developing conceptual ABSS models. *We propose a tool that utilises Conversational AI (CAI) and the EABSS framework to generate innovative conceptual ABSS models in a short amount of time and with a minimum of upfront case-based knowledge required.* The use of the EABSS framework ensures a comprehensive and well-defined conceptual ABSS model tailored to its intended purpose. This specification encompasses not only the details of the model design but also a complete experimental framework for the intended simulation and analysis.

The contribution this paper makes to the field of ABSS is the establishment of a novel paradigm for generating conceptual ABSS models through the successful integration of CAIS with the EABSS framework, providing evidence that automated assistance can effectively support the traditionally manual process of developing conceptual ABSS models. We present a generic prompt script, which leverages the iterative information retrieval mechanisms embedded in the EABSS framework to support mimicking on the fly fine-tuning the Large Language Model (LLM) that drives the CAIS. This script can be used for any ABSS case study with only minor modifications. To illustrate the prompt script application and evaluate the proficiency and coherence of the generated conceptual ABSS model, we present a proof-of-concept application of the prompt script to generate such a specification for a case study on the impact of adaptive architecture in a museum environment.

## 2 BACKGROUND

### 2.1 Agent-Based Modelling and Agent-Based Social Simulation

Agent-Based Modelling (ABM) is a modelling approach that focuses on describing a system from the perspective of its constituent units, known as agents, and their interactions (Bonabeau 2002). These agents are designed to mimic the behaviour of their real-world counterparts (Twomey and Cadman 2002). ABSS is a specific subset of Social Simulation that uses ABM to model individuals and their interactions within social structures. It specifically aims to identify social emergent phenomena, such as social cohesion, social norms and cooperation, spatial patterns, or cultural dynamics. A comprehensive analysis of the topic can be found in Dilaver and Gilbert (2023).

## 2.2 Co-Creation

Co-creation is a collaborative process, promoting the active and equal participation of multiple stakeholders (in our case including domain experts, researchers, and end-users) in the generation, development, and refinement of ideas, products, or services (Ind and Coates 2013). It is characterised by a shared decision-making approach, wherein diverse perspectives and expertise are integrated to foster innovation and problem-solving (Prahalad and Ramaswamy 2004). As stated by Chesbrough (2011) the co-creation approach emphasises the active involvement of stakeholders throughout the entire innovation process, ensuring a more contextually relevant and user-centred outcome.

The use of co-creation in ABSS is particularly beneficial in addressing complex social phenomena. By involving stakeholders in the conceptualisation and parameterisation of agent behaviours, the resulting models capture a broader range of factors influencing social systems (Smetschka and Gaube 2020). Hence, such a participatory approach contributes to the development of more contextually relevant and realistic simulations.

## 2.3 Engineering Agent-Based Social Simulations

The EABSS framework is a structured and transparent approach that applies best practices from Software Engineering to support various use cases related to ABSS modelling, such as model conceptualisation, model documentation, reverse engineering of model documentation, discussions to analyse research topics, and blue sky thinking for defining novel and innovative research directions (Siebers and Klügl 2017). Figure 2.1 shows a high-level overview of the framework. Adhering to the framework's prescribed process flow results in a comprehensive and transparent conceptual ABSS model tailored to the intended purpose. This specification encompasses not only the details of the model design but also a complete experimental framework for the intended simulation and analysis. While the EABSS framework shares similarities with tools like the ODD protocol (Grimm et al 2006), a protocol designed for ABSS documentation, the EABSS framework emphasises the actual model development process.

The framework structure is based on the design principle of "separation of concerns" (Dijkstra 1974) commonly used in Software Engineering, which explains the split of the process flow into Analysis and Design and the embedded steps, each focussing on a specific concern within the system conceptualisation. The Analysis steps support defining the problem and capturing the study requirements, identifying user needs and system constraints. Analysis activities involve gathering information through stakeholder interviews, surveys, and existing documentation to establish a clear understanding of what the system must achieve. The Design steps focus on creating the architecture and components of the system to meet the specified requirements. Design activities include defining the system's structure, interfaces, and interactions to ensure that it is efficient, maintainable, and scalable.

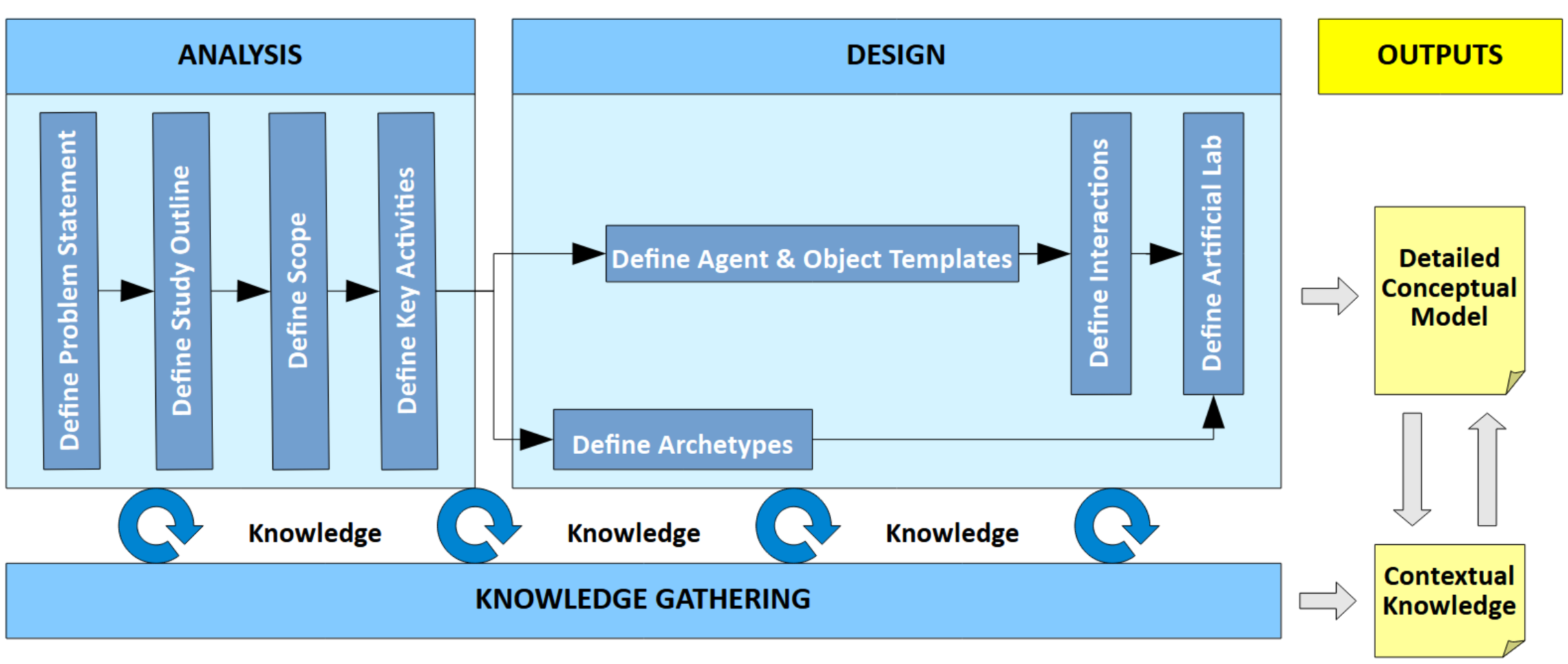

Figure 2.1 Overview of the EABSS framework (adapted from Siebers and Klügl 2017)

While the general workflow is inherited from Software Engineering, the individual steps within the workflow are adapted and specific to fit the requirements for conceptual ABSS modelling. In addition we have added one component, knowledge gathering, to make its process flow explicit. During the conceptual model development participant knowledge provides the foundation for building the conceptual model step-by-step. In addition, during this process contextual knowledge is generated that does not directly relate to the individual steps. This includes for example stakeholders' agreement on terminology or stakeholders personal experience with certain situations or artefacts. Contextual knowledge can later be used as one resource for validating the conceptual model.

The EABSS process is grounded in the concept of co-creation (Mitleton-Kelly 2003) and ideas from Software Engineering (Sommerville 2015). In addition, it draws on elements of Kankainen's focus group approach to service design (Kankainen et al 2012). It implicitly provides ground rules, which is something commonly done when working with children but often forgotten when working with adults. These ground rules are in line with De Bono's philosophy of parallel thinking (De Bono 1985), and state that people are going to listen to each other, and that people respect each other's point of view. A focus group format is used for running EABSS sessions. To capture information the framework uses several predefined interactive tasks, table templates, and the Unified Modelling Language (UML), which is a graphical notation commonly used in Software Engineering, as main forms of stimulating and documenting contributions from all participants during the co-creation process. In the following we provide a brief description of the purpose of each individual step within the EABSS framework:

**Step 1**: Define the Problem Statement. In this step, we formulate a comprehensive statement that articulates the study's general purpose, including relevant information detailing the study's title, domain, approach, context and aim.

**Step 2**: Define the Study Outline. Here, we establish an experimental framework by specifying objectives and constraints to be fulfilled and/or hypotheses to be tested. In addition we define a list of experimental factors (parameters) to allow creating scenarios relevant to testing objectives and/or hypotheses and a list of responses (outputs/statistics) for measuring if objectives have been achieved and/or to test if hypotheses should be accepted/rejected.

**Step 3**: Define the Scope. This step involves deriving a list of entities (key actors and elements of the physical environment) and concepts (social and psychological aspects) potentially relevant for the ABSS and then deciding which of these elements to include in the conceptual ABSS model to be created.

**Step 4**: Define Key Activities. In this step, we capture potential interactions between multiple key actors and between key actors and the physical environment by assigning key actors to relevant actions or activities (also known as use cases).

**Step 5**: Define Archetypes. Here, we focus on categorising key actors into archetypes that will later serve as agents in the simulation. Using categorisation schemata, we separate a synthetic population into behaviourally distinct groups, where each archetype represents a fundamental social role or personality.

**Step 6**: Define Agent and Object Templates. In this step, we create templates to represent agents (such as individuals, groups, or organisations), and objects (elements of the physical environment). With the help of diagrams we capture structural design and relationships between entities (class diagrams), possible states of an entity and the events or the messages that cause a transition from one state to another and the action that results from a state change (state charts), and how activities are co-ordinated (flow charts).

**Step 7**: Define Interactions. Here, we use sequence diagrams to capture the flow of interactions between agents and between an agent and the physical environment. These diagrams provide a clear visual representation of the order and structure of such interactions. This is especially important in scenarios where the timing or sequence of interactions can impact the model's outcomes.

**Step 8**: Define the Artificial Lab. In this step, we design an artificial lab environment in which we can embed all our entities and define some global functionality. We need to consider things like global variables (e.g. to collect statistics), compound variables (e.g. to store a collection of agents and objects), and global functions (e.g. to read/write to a file). We also need to make sure that we have all variables in place to set the experimental factors and to collect the responses we require for measuring if objectives have been achieved and/or to test if hypotheses should be accepted/rejected.

The EABSS framework has been tested for all the potential use cases mentioned above (model conceptualisation, model documentation, etc.) in different domains, including Architecture, Geography, Organisational Cognition, Service Management, and Digital Mental Health. It uses a step-by-step approach designed to look at a complex system in more detail with every further step. Also, there is always information from previous steps that can be reused to initiate the next step. Reusing previous information in most cases avoids getting stuck during the modelling process. Still encountering difficulties is a clear indication that something in the preceding steps is incorrect and requires modification. Hence, the framework includes a built-in validation mechanism.

The outcome of an EABSS co-creation session is a structured record of the key points of the focus group discussions, in a format that is easy to understand by all stakeholders as well as the people tasked with implementing the model, and easy to extend. With a little effort, this can be implemented as a simulation model, allowing to conduct simulation experiments. Academic examples that demonstrate the use of the framework can be found in Siebers and Klügl (2017), Barnes and Siebers (2020), and Mashuk et al (2024).

### 2.4 Conversational AI

The term 'conversational AI' refers to the concept of AI that can engage in natural dialogue. The term 'conversational AI system' refers to an implementation that uses conversational AI technology to engage in natural language conversations with users. Such systems, often powered by Natural Language Processing (NLP) and Natural Language Generation (NLG) models, are computer programmes designed to engage in conversation with users by understanding user inputs, processing text to discern meaning, context, and intent, and subsequently generating human-like language responses to facilitate interactions and provide assistance in various applications (Adamopoulou and Moussiades 2020; Crothers et al 2023). The vast majority of current CAISs use neural language models based on the Transformer Architecture proposed by Vaswani et al (2017). Neural language models use deep neural networks to understand and generate human-like language. They are pre-trained on vast amounts of diverse text data, enabling them to capture intricate patterns and nuances of language. Additionally, they often incorporate machine learning techniques, allowing them to improve their performance over time by learning from user interactions and feedback. Large Language Models (LLMs) like GPT-3 are a specific class of neural language models, characterised by their vast size and training data. ChatGPT is a prominent example of a CAIS that uses the GPT-3 architecture and integrates both, NLP and NLG, leveraging pre-training and machine learning to generate human-like responses based on the context provided in the conversation (Nazir and Wang 2023). The underlying mechanism of ChatGPT involves generating responses by predicting the next word or sequence of words given the context of the conversation. This process is driven by the model's ability to understand and generate coherent language based on the patterns it has learnt during training (Briganti 2023).

At the time of writing, ChatGPT, developed by OpenAI, is the most prominent CAIS. Others include BERT, T5, and Gemini developed by Google, ERNIE, developed by Baidu, BART, developed by Facebook AI Research, and Llama, developed by Meta AI. In-depth information about these and many other systems and models that have been developed since can be found on the Hugging Face Machine Learning and Data Science platform (Hugging Face 2024a).

While CAISs like ChatGPT showcase the huge potential of AI in NLP and NLG, it is essential to acknowledge their limitations, including their struggle to deal with ambiguity, the lack of real-world understanding, and the possibility of generating incorrect or biased responses (Hadi et al 2023). Therefore, they should be used as supplementary tools to support the work of humans, rather than relying solely on such tools for tasks that require nuanced comprehension, contextual awareness, and a comprehensive understanding of complex, real-world scenarios.

## 2.5 Related Work

Literature on ABSS in conjunction with terms related to CAISs (Large Language Model, Generative AI, Conversational AI, and ChatGPT) is still very sparse. Several papers explored the technical aspects and issues of integrating LLMs into ABMs to enhance the capabilities of simulated agents in terms of realistic human-like decision-making, social interactions, and communication within diverse simulated environments. Examples include Vezhnevets et al (2023), Wu et al (2023), and Chen and Wilensky (2023). All of these papers reached a consensus that LLM have significant potential for enhancing simulation research. However, more research is needed to fully understand their capabilities. Other papers focused on Generative ABM (GABM) and its application in understanding social system dynamics. GABMs are ABMs in which each agent is connected to an LLM. This allows agent reasoning and decision-making to be delegated to the LLM, which can perform a qualitative analysis before making decisions. As a result, the simulation exhibits more nuanced and realistic behaviours (Ghaffarzadegan et al 2024). This is a significant advantage over traditional ABMs, which rely on predefined rules and parameters. GABM has been used to explore scenarios such as social norm diffusion (Ghaffarzadegan et al 2024), epidemic modelling (Williams et al 2023), and opinion dynamics (Chuang et al 2023), aiming to capture complex social phenomena. As before, authors were very optimistic about the new technology, but also stated that there are numerous obstacles that need to be addressed. A more optimistic view is adopted by Wang et al (2024). They model individuals within an urban environment as LLM agents using a novel LLM agent framework for personal mobility generation. This framework employs an activity generation scheme that directs the LLM to simulate a specific individual based on a predetermined activity pattern. Subsequently, it meticulously generates an activity trajectory that aligns with the individual's daily motivations. The authors claim that they have successfully demonstrated the potential of LLMs to enhance urban mobility management.

Literature on Co-Creation in conjunction with the terms related to CAISs exist, but on closer inspection the papers found all use the term co-creation to describe the interaction between humans and CAISs, rather than the concept of stakeholder discussion and shared decision-making. We did not find any papers that link Participatory Modelling to the terms related to CAISs.

Finally, we looked at the terms Conceptual Modelling and Conceptual Model in conjunction with the terms related to CAISs. While most papers in this category focused on systems engineering, they offer valuable insights that align well with the objectives of our research initiative. Noteworthy amongst these papers is the work of Härer (2023) who developed a conceptual model interpreter that uses an iterative modelling process, generating and rendering the concrete syntax of a model (in the form of class diagrams and directed graphs) conversationally. The results indicate that modelling iteratively in a CAIS is a viable option. In contrast, Giabbanelli (2023) discusses technical concepts of mapping each of the core steps in a simulation study life cycle (conceptual modelling; simulation experiment; simulation output; verification and validation) to a task in NLG. The paper concludes that it would be exciting if it would be possible to automatically generate a functional simulation model prototype from a conceptual model created with the help of a CAIS, but that the current state of the art is not advanced enough to make this happen.

In conclusion, our analysis of related work reveals a significant gap in exploring the integration of CAISs into workflows related to conceptual modelling in ABSS. We only found one closely related paper, written by Giabbanelli (2023), which approaches the topic from a methodological perspective, providing in-depth guidance for the use of different technologies to achieve the goal of holistic simulation study automation, without giving a concrete implementation example. In contrast, we will focus on the practical side by demonstrating the capabilities of CAISs in supporting the process of co-creating conceptual ABSS models. The successful use of LLMs in GABM demonstrates their potential in supporting conversations and decision-making, both of which are critical aspects of the co-creation process we intend to implement.

# 3 METHOD

One can consider multiple scenarios for incorporating CAI support into the conceptual ABSS modelling process, utilising the EABSS framework as its basis: (1) a CAIS as an idea generator for developing novel and innovative project ideas, (2) a CAIS as an impersonator for representing absent focus group participants, providing their perspectives during discussions and debates, (3) a CAIS as a trainer for EABSS moderators, (4) a CAIS as a mentor, helping when ideas are scarce, before, during, or after each EABSS step within a focus group session, and (5) a CAIS as an expert for filling gaps in a focus group report after the focus group session is completed.

In our research activities, we focus on the first two scenarios: the use of CAI support for idea generation and the use of CAI support for enacting missing focus group participants. Rather than relying on real-world focus groups, we explore the potential of using CAISs to develop a comprehensive conceptual models for ABSS projects. To generate such a specification with a CAIS, it is essential to adopt a structured approach, as the sequence in which information is gathered plays a crucial role in information management and retention. Given the current limitations in LLM memory, we also need a strategy to retain relevant information derived in earlier stages of model development to effectively support subsequent steps. In both cases the EABSS framework can provide valuable assistance.

### 3.1 Prompt Engineering

Currently, the craft of prompt engineering is more an art than a science. However, there are well-established general principles that prove helpful for creating draft prompts, which can subsequently be fine-tuned for a specific purpose or application domain. Busch et al (2023) state that a standard template for a prompt should adhere to the following structure: "context or background information, clear instructions, example(s), constraints or guidelines, acknowledgement of ambiguity, and feedback loop". To illustrate this approach, we have crafted the following prompt, aimed at generating an explanation of Einstein's Theory of Relativity: "*Take on the role of a senior physicist writing the background section of a journal article on Albert Einstein's Theory of Relativity. Explore its significance in Physics, delve into the historical context, and analyse its impact on our comprehension of space, time, and gravity. Your essay should encompass an overview of the key principles, such as the equivalence of mass and energy, time dilation, and the curvature of spacetime. Provide concrete examples, such as the influential thought experiment with a moving train. Adhere to a 500-word limit, ensuring a balanced blend of technical details accessible to a broad audience. Support your information with citations and acknowledge potential ambiguities, offering your perspective on multiple interpretations. Use the feedback loop to evaluate the accuracy, clarity, and coherence of the essay, emphasising the need for engaging and informative content in any revisions*". This prompt adheres to the structure mentioned above. Of course, it is not always necessary or even advisable to strictly follow this pattern, but it provides a good foundation for prompt engineering.

### 3.2 Scripting Principles

We decided to present the collection of prompts needed to generate a conceptual ABSS model as a holistic script that follows certain guiding principles: (1) it needs to be easy to use, containing prompts that are based on the standard prompt template defined in Section 3.1, whenever possible, (2) it needs to be accessible for everyone, i.e. it should work with the free ChatGPT 3.5 version, (3) the user needs to have a choice between execution speed and response quality, (4) it needs to be able to provide tables and diagrams as responses, as these are essential components of the EABSS framework, (5) its chained prompts need to follow some design patterns, for transparency, to prevent errors, and to make it easy to maintain and extend, (6) it needs to support user interaction, e.g. for revising responses, (7) it needs to generate an informative report, in line with the EABSS framework output, and (8) it needs to provide some kind of language support, to make sure everyone understands the terminology used within the report. In the following, we will refer to this script as the "EABSS script".

### 3.3 Embedding the Concept of Co-Creation

Co-creation is a valuable tool for fostering innovation, generating creative solutions, and building shared understanding among diverse stakeholders (Loureiro et al 2020). A CAIS cannot replicate real-world co-creation sessions, as it lacks the ability to engage in critical reflection, challenge assumptions, and apply logical thinking to analyse and synthesise ideas. In genuine co-creation, participants contribute unique, subjective perspectives grounded in lived experiences, creating a dynamic interplay between intuition and structured reasoning (Kimmel and Hristova 2021). While a CAIS can simulate reflection by drawing from patterns in its training data, it falls short of the depth and spontaneity of human critical thinking, essential for authentic co-creation processes.

However, there are benefits in adding the notion of co-creation to the EABSS script. While the discussion is very stilted and clearly artificial, it helps the AI to consider the viewpoints of different stakeholders when generating subsequent responses, reducing the potential bias introduced by considering only the viewpoint of a "general" stakeholder. In addition, the AI's ability to generate topic-related random questions during the simulation of a focus group session can lead to interesting co-creation responses that we might not have considered otherwise, thereby providing valuable contextual knowledge. This is particularly evident as each execution of the EABSS script generates a different set of questions.

### 3.4 Controlling Conversation Flow through Human Intervention

Standard conversations with CAISs involve a continuous exchange between the CAIS and the user, where the user's input influences the responses and directs the AI's focus (Menon and Shilpa 2023). This can limit creativity, as the CAIS is prone to the user's mental framework. This 'human-in-the-loop' approach is suitable when the user has a specific design in mind and only requires assistance with details. The EABSS script allows for course corrections throughout the conversation after each prompt submission, enabling the user to steer the conversation by requesting the CAIS to add, remove, or modify certain artefacts or to reconsider its previous responses. However, when the user seeks truly creative solutions (blue-sky thinking) aiming to use the CAIS as an independent model designer, it may be more effective to minimise interference and avoid directing the CAIS too much.

For our proof-of-concept application of the EABSS script, we aim to demonstrate the CAIS's ability to work independently, with guidance provided only through the application of the EABSS framework and a brief description of the case study context, in order to assess the capabilities of the CAIS rather than those of the modeler.

### 3.5 Controlling the Level of Stochasticity for LLM responses

There are two methods for controlling the level of stochasticity in LLM responses: defining it through parameters (Temperature and Top_p) or through prompt engineering (by instructing the LLM to imitate specific parameter settings or by incorporating controlling commands directly into the prompt).

In the first approach, the balance between creativity and relevance in the LLM's responses is managed by adjusting the "Temperature" and "Top_p" parameters (Nguyen et al 2024). Temperature governs the randomness of the output, with lower values (closer to 0) making the responses more deterministic and repetitive, and higher values (closer to 2) promoting greater diversity and randomness. Top_p (nucleus sampling) influences the cumulative probability distribution of the LLM's predictions, effectively filtering out less probable options. A lower Top_p value results in a more focused LLM that prioritises high-probability outcomes, thereby reducing randomness, while a higher Top_p allows for a broader range of possibilities, increasing diversity in the generated text. As many CAISs, including ChatGPT, do not provide access to these parameters through their web interface, these parameters can only be controlled via their API, which typically requires a paid subscription.

This limitation brings us to the second option, which involves influencing the LLM's behaviour by instructing the CAIS to imitate different parameter settings, a similar tactic to imitating personas (Thominet et al 2024). This results in responses that approximate the expected outputs from specific parameter adjustments through the API. However, this method only provides an approximation and requires some trial and error to achieve the desired level of stochasticity. To illustrate, let us reconsider Einstein's theory of relativity. The prompt >>*Imitate a VERY LOW TEMPERATURE setting (low randomness compared to your default settings) and a VERY HIGH TOP_P setting (high coherence of response compared to your default settings). Describe Einstein's theory of relativity in one sentence.*<< generated the response: "*Einstein's theory of relativity states that space and time are interwoven into a single continuum, and that the presence* *of mass and energy warps this fabric, affecting the motion of objects and the passage of time.*" In contrast, the prompt >>*Imitate a VERY HIGH TEMPERATURE setting (high randomness compared to your default settings) and a VERY LOW TOP_P setting (low coherence of response compared to your default settings). Describe Einstein's theory of relativity in one sentence.*<< generated a more creative but less relevant response: "*Einstein's relativity is like a trampoline made of starlight, where planets bounce, time melts like jelly, and the speed of light is a neon snack!*". Alternatively, "controlling" commands can be directly incorporated into the prompt. For instance, the prompt >>*Be SCIENTIFIC. Describe Einstein's theory of relativity in one sentence.*<< generated the response: "*Einstein's theory of relativity, comprising both the special and general theories, posits that the laws of physics are invariant across inertial frames of reference, and that gravity is the result of spacetime curvature caused by mass and energy.*" Conversely, the prompt >>*Be CREATIVE. Describe Einstein's theory of relativity in one sentence.*<< generated the response: "*Einstein's theory of relativity is the universe's secret dance, where time pirouettes and space sways, pulling and pushing on one another, revealing that nothing stands still - not even light - because everything is caught in the graceful, inevitable curve of gravity's embrace.*"

For the purposes of this study, we found that a combination of both of these prompt engineering techniques is the most effective approach. Specifically, setting an overall tone for the responses at the beginning of the conversation by imitating parameter settings, and then requesting creativity and innovation in specific instances through prompt engineering, while clearly defining the boundaries for such creativity, and subsequently returning to the initially established tone.

### 3.6 Large Language Model Fine-Tuning

Fine-tuning is a way to take a pre-trained LLM, which is good at general language understanding, and make it a specialist in a particular area. This requires a dataset of text that is relevant to the specific task the LLM should perform. With a smaller, task-specific dataset, one essentially adjusts the LLM's internal parameters to make it better on the chosen task (Ozdemir 2023).

Besides the purpose of providing a structured approach to conceptual ABSS modelling the EABSS framework also supports mimicking the fine-tuning process of the LLM underlying ChatGPT for each specific case study. As previously stated, in the EABSS framework, what is derived in one step serves as the foundation for the next. This mechanism supports the learning of the LLM for a specific case on the fly, hence training it for the specific task at hand, without any additional effort. It mimics the active few-shot fine-tuning using transductive active learning (Hübotter et al 2024). Only relevant information is kept, which makes this approach very efficient.

### 3.7 Concept Validation

To assess the proficiency of the EABSS script, we employ a case study approach, using one of our previous studies, specifically "Adaptive Architecture in a Museum Context" (Siebers et al 2018). Our motivation for selecting this study is that it presents a topic requiring an innovative solution since this idea is novel. Hence, the LLM has less training data on it compared to classical models such as predator-prey and segregation, making it more challenging for the LLM to generate a coherent storyline. Furthermore, we were curious to see what solution the LLM would generate, given the challenges we faced in formulating a solution during our original study.

When applying the EABSS script, the only information we provide to ChatGPT in relation to this case study is a broad summary of the relevant topic based on ideas from our previous attempt and an indication of the related domain. All the remaining information is generated by the CAIS. For this reason, our expectation is that we will get a conceptual model that focuses on different aspects, and our hope is that it will provide some innovative ideas that could be useful for further development or practical application.

For the assessment we decided to use evaluation by human experts (Botpress Community 2024). While there are more formal methods for evaluating the quality of CAIS responses, such as Perplexity (Hugging Face 2024b), these methods are intended for measuring the impact of training on the underlying LLM (Hugging Face 2024c) rather than evaluating the quality of prompts. Therefore, we have created a set of indicators to assess whether the novel CAI supported approach to conceptual model development produces substantive output. These comprise: (1) Usability: Is the EABSS script easy to use? (2) Generality: Is the script tailored to a specific problem, or is it more general-purpose? (3) Pertinency: Did the CAIS produce an output report relevant to the provided topic? (4) Readability: Does the output generated by the CAIS exhibit coherence and flow, facilitating reader comprehension? (5) Conformity: Does the CAIS's resulting report appropriately cover all aspects of the EABSS framework? (6) Believability: Do the responses generated by the CAIS effectively mimic human responses? (7) Originality: Does the storyline generated by the CAIS, based on its unique interpretations of a specific topic, embody fresh and innovative concepts?

## 4 EABSS SCRIPT DETAILS

### 4.1 EABSS Script Foundation

The script we created is based on the EABSS framework presented in Section 2.3. It considers each of the eight steps inherent in the framework. Information reuse guidance is embedded in the form of requirements. While the script is closely related to the EABSS framework, it is not an exact match. To fully use the capabilities of ChatGPT, we added some additional features to our script that are easy to generate and support the knowledge build-up during the conceptual modelling process, providing some additional contextual insight. An example of such an addition is the definition of terminology.

### 4.2 EABSS Script Segmentation

The script is separated into four main segments. Segment 1 provides instructions to ChatGPT concerning the style of responses and encourages ChatGPT to generate high-quality output, based on the chat history. This relates to the iterative information retrieval concept embedded in the EABSS framework. Segment 2 focuses on the analysis of the problem to be investigated, defining a problem statement, study outline, model scope, and key activities. This relates to Steps 1-4 in the EABSS framework. Segment 3 covers the model design, defining archetypes, agent and object templates, interactions, and the artificial lab. This relates to Steps 5-8 in the EABSS framework. Segment 4 provides some concluding remarks, testifying that the aim has been achieved, answering the questions related to the objectives and hypotheses, and listing identified limitations and ideas for overcoming these limitations through further work.

### 4.3 EABSS Script Implementation Strategies and Patterns

As stated in Section 3.1, prompt engineering is a challenging task. Prompts can easily get messy and unproductive. One can quickly lose the overview when working on complex prompts that depend on each other and should be executable in a chain. In light of this, our objective during the scripting process was to construct a well-structured script that includes clear prompts, with the added benefit of making it easy to maintain and extend the script. To achieve this goal, we established a series of prompt design strategies and patterns that we consistently applied during the development of the EABSS script.

#### 4.3.1 Prompt design strategies

The web is full of suggestions related to prompt engineering and several books have been published on the topic, such as Sarrion (2023). However, consensus is lacking among these resources, and reliable documentation or well-established standards are not available. In Section 3.1 we have outlined the main principles for crafting effective prompts. Here, we elaborate on implementing these principles by presenting a collection of prompt design strategies and providing examples of how they are used in conjunction with the EABSS script.

**Chat preparation**: It is important to prepare ChatGPT for the job at hand. This can be done by providing some initial commands to improve ChatGPT's memory skills, response clarity, and quality. In addition, selecting the appropriate values for Temperature and Top_p is crucial for balancing creativity and coherence, as explained in Section 3.5. In conceptual modelling, we want to foster a flow of ideas that are diverse yet still logically connected. Furthermore, we need flexibility in generating different perspectives, but without sacrificing the structure and relevance of the content.

The following prompt could be used for this purpose: >>*You are ChatGPT, a language model developed by OpenAI. Consider the ENTIRE conversation history to provide 'accurate and coherent responses'. Imitate a MEDIUM TEMPERATURE setting of 0.9 (for a creative yet structured approach, encouraging new ideas without losing coherence) and a VERY HIGH TOP_P setting of 0.9 (promoting diversity in the responses while ensuring logical connections within the generated content). Use clear, precise language during the entire conversation. Prioritise substance during the entire conversation.*<<. If a chain of individual prompts is used, it is recommended to instruct ChatGPT to work through them step-by-step. The following prompt could be used for this purpose: >>*Step-by-step, work through the following task list in the given order during the entire conversation.*<<. It should be placed at the beginning of a list of bullet point prompts.

**Avoiding response display**: ChatGPT can be directed not to display a response on the screen for some time. The response will still be generated but will not be displayed. This reduces the required waiting time between prompt submissions, in particular when prompts generate verbose responses. The following prompt could be used for this purpose: >>*Do not print anything! Got it? Say yes or say no!*<<. ChatGPT will respond to this with either a "Yes" if all prompts pasted in one shot were executed successfully, or with a "No" if there were issues. It is important to place the output-avoiding prompt at the end of the content pasted into ChatGPT in one shot, which can be a single prompt or a collection of prompts of any size.

**Roles and tones**: It is recommended to define roles and tones for ChatGPT to adopt during a conversation, as these have a significant impact on the content and style of its responses. The first decision to make involves setting the defaults for the entire conversation. The following prompt provides an example: >>*Take on the "role" of a "Sociologist" with experience in "Agent-Based Social Simulation" during the entire conversation, unless instructed otherwise. Use a "scientific tone" during the entire conversation unless instructed otherwise.*<<. During the chat, role and tone can be temporarily changed, either jointly or independently from each other. The following prompts provide an example for changing both: >>*Take on the role of a "Focus Group Moderator" with experience in "architecture". Use a "debating tone".*<< … >>*Get back to the previous role. Get back to the previous tone.*<<.

**Reuse of information**: All LLMs are stateless by design, which means that the LLM itself does not retain any conversation information. Instead, the context is retained in the ChatGPT web application in the form of the actual conversation (Kelk 2023). What happens behind the scenes during a conversation is that when ChatGPT receives a new prompt, it re-reads the entire conversation. Hence it appears that ChatGPT has a memory. Once a conversation gets too long (the maximum for ChatGPT 3.5 is around 3000 words in a single conversation) ChatGPT removes pieces of the conversation from the beginning at the same level as new information is added to the end. Hence it appears that ChatGPT forgets information.

To minimise information loss during a conversation, it is essential to create prompts that hold only necessary commands and are formulated concisely. In this context it is recommended to carefully evaluate and improve prompts, if necessary, to ensure they satisfy those requirements. Additionally, commands should be embedded into these prompts to ensure only relevant information is provided concisely, avoiding repetition and meandering.

However, when we consider that we are attempting to imitate an EABSS focus group discussion that lasts for several hours, it could be argued that ChatGPT's forgetfulness is not entirely negative. In reality, participants often forget facts during such long focus group discussions. If needed, information can be retrieved through "memorised keys", which represent the notes taken during a real-world focus group session. The following prompt provides an example: >>*Define the "aim" for the study in 40 WORDS (if possible). Memorise this aim as {key-aim}. List the memorised key-aim.*<<. As soon as these memorised keys are used, they are refreshed in ChatGPT's memory. Hence, if there are concerns about losing valuable information, the relevant memorised keys could be accessed, extending their retention time in ChatGPT's memory.

**Prompt clarity and structure**: Using symbols or characters in prompts can enhance clarity and structure. Unfortunately, there are no official guidelines on this topic, so we decided to ask ChatGPT itself for some guidelines. After intensive testing, the ones we found most useful for our script are bullet points (to list items in a clear and organised manner), numbered lists (to present information in a specific order), colons (to introduce a list, explanation, or elaboration), quotation marks (to indicate that a specific term or phrase should be used), squared brackets (to provide additional information or clarify a point), ellipses (punctuation mark (…), to indicate a continuation or omission), parentheses (for additional comments or information that is related but not crucial to the main prompt), slashes (to present two options or alternatives), curly braces (to represent choices, variables, or placeholders), capitalisation or bold text using **…** (to emphasise importance), and vertical bar between command sequences (to imitate a Unix pipe).

**Script-based visualisation**: ChatGPT 3.5 cannot generate diagrams directly. However, it can generate scripts for applications that utilise them to generate diagrams. Examples of applications that are capable of creating UML diagrams (the diagram type used by the EABSS framework) from scripts are PlantUML (PlantUML 2024) and Mermaid.js (Mermaid 2024). Unfortunately, generated scripts often contain errors. Therefore, it is important to take extra care when creating visualisation-related prompts, to ensure that reoccurring errors are dealt with by the prompt itself.

Definition of relevant terminology: It is advisable to ask ChatGPT to define relevant terms to ensure that the user's understanding aligns with ChatGPT's interpretation. Here, ChatGPT must act in the correct role for the context, as its interpretation of terminology may vary depending on the specific domain or context. In addition, this practice supports users not familiar with all the relevant terminology used in the EABSS framework.

**Refining responses**: Sometimes, ChatGPT provides responses that lack sufficient depth or omit required elements, features, or options. In such cases, it is necessary to refine the response through course correction using a human-in-the-loop approach. While nothing can be deleted from a conversation, memorised artefacts in keys can be updated. When referring to them later, the updated content will be considered, instead of the original one. For the refinement, it is helpful to have a library of alteration and reinforcement prompts. The following prompts are examples of tweaking responses (replace "x" with the relevant content): *>>Remove x. Update the memorised key.<<*; *>>Add x. Update the memorised key.<<*; *>>Increase complexity. Update the memorised key.<<*; *>>Critically reflect and improve x based on your reflection. Update the memorised key.<<*.

### 4.3.2 Prompt Design Patterns

In Software Engineering, design patterns represent high-level solutions to common problems that occur during software design. They tackle recurring design challenges by offering general guidelines and solutions derived from best practices that have evolved and are rooted in well-established design principles. This enables the creation of code that is more modular, maintainable, and scalable. Building upon this concept, we have created four design patterns that are utilised in the EABSS script. While these patterns are not intended to be rigorously enforced, they offer guidance and enhance transparency in prompt design. In the following, we present the commands embedded in our design patterns in the form of command chains. Commands labelled "OPTIONAL" are not relevant for all command chain implementations that are based on the related pattern. Commands labelled "INTERVENE" are opportunities for the script user to influence the responses of ChatGPT manually. Text within curly braces and the curly braces themselves have to be replaced with the details mentioned in these braces.

**General pattern commands**: (1) Take on the role of {provide a role related to the specific activity of the step} with experience in {provide an experience related to the specific activity of the step}. (2) Provide definitions of relevant terms in the context of the role adopted: {provide a list of terms to be defined}. (3) Define several relevant elements for the step. (4) OPTIONAL: The following requirements must be satisfied when choosing these elements: {provide a numbered list of requirements}. (5) Provide further specifications ({provide a list of comma-separated specification types}) for these elements. (6) OPTIONAL: Use provided output format: {provide the desired output format} (7) Memorise details using a {provide key placeholder name}. (8) INTERVENE: Add (or remove or change) {provide relevant element (and statement of action)} and update related memorised {provide related key placeholder}. (9) INTERVENE: Increase complexity and update related memorised {provide related key placeholder}.

**Co-creation pattern commands**: (1) Play a co-creation role-play game in which all the memorised key-stakeholders discuss with each other potential {provide topic to be discussed} for the study considering the pros and cons. (2) Use a "debating tone". (3) The moderator focuses on 1 novel RANDOM question. (4) Provide the question and the details of the controversial discussion. (5) Agree on {add required number} potential {provide topic under discussion} that satisfy the view of all participating memorised key-stakeholders. Memorise these potential {provide topic under discussion} as {provide key placeholder name}. (6) OPTIONAL: Propose {add required number} criteria for ranking the {add required number} potential {provide topic previously discussed} to support the decision which {provide topic previously discussed} to carry forward. (7) OPTIONAL: Use provided output format: {provide the desired format}. (8) Use a "scientific tone".

**Table pattern commands**: (1) Use provided table format and creation rules: {provide table format and creation rules}. (2) Define {provide several relevant elements for the step}. (3) OPTIONAL: The following requirements must be satisfied when choosing these elements: {provide a numbered list of requirements}. (4) Provide further specifications ({provide a list of comma-separated specification types}) for these elements. (5) Follow the guidance on how the information should be organised within the table: {provide a list of guidelines}. (6) Memorise details using a {provide key placeholder name}. (7) INTERVENE: Add (or remove or change) {provide relevant element (and statement of action)} and update related memorised {provide related key placeholder}. (8) INTERVENE: Increase complexity and update related memorised {provide related key placeholder}.

**Diagram pattern commands**: (1) Generate a script for a 'comprehensive {provide a specific UML diagram type} diagram' in "Mermaid.js". (2) Use the following information for the diagram design: {either provide a memorised key or a numbered list of required information types}. (3) The following requirements must be satisfied when creating the diagram: {provide a numbered list of requirements (often requests to provide specific elements of a diagram or to avoid systematic errors in a diagram; for the latter, it is helpful to provide examples of correct solutions)}. (4) OPTIONAL: Increase complexity and add additional features: {comma-separated list of additional features}. (5) OPTIONAL: Critically reflect and improve the script based on your reflection. (6) Memorise details using a {provide key placeholder name}. (7) INTERVENE: Add (or remove or change) {provide relevant element (and statement of action)} and update related memorised {provide related key placeholder}. (8) INTERVENE: Increase complexity and update related memorised {provide related key placeholder}.

## 5 VALIDATION EXPERIMENT

As stated in Section 3.7, we use a case study approach to assess the proficiency of the EABSS script, using one of our previous studies, specifically "Adaptive Architecture in a Museum Context" (Siebers et al 2018). The goal of the original case study was to generate ideas for using adaptive architecture in a futuristic museum within an exhibition room that is visited by adults and children. The adaptive architecture we planned to integrate consisted of large wall-mounted screens on which smart content windows were moving with the visitors. During our focus group discussions, we identified another artefact of interest, a smart partition wall that creates a dynamic and flexible exhibition space by constantly analysing visitor movement and physically relocating itself. Through real-time decision making, the partition wall optimises the floor space for the visitors located in each of the sections it created.

We prepared the EABSS script by providing a broad summary of the relevant topic and an indication of the related domain. It is important to note that this validation experiment aims to test whether the CAIS can assist in generating a feasible storyline, resulting in an understandable, meaningful, and cohesive conceptual ABSS model. Our goal is not to replicate the existing conceptual model from the original case study. We have selected a rare case with limited training data for the LLM to demonstrate that the CAIS can handle more challenging scenarios. Given how LLMs function, we can confidently expect that more common use cases will yield even better results. For the assessment, we used the criteria defined in Section 3.7: usability, generality, pertinency, readability, conformity, believability, and originality.

In this section we will only present samples of the prompts we supplied to the CAIS during the validation experiment along with the corresponding responses it generated. The complete EABSS script, encompassing all prompts used during the validation experiment, can be found in Appendix 1. The complete validation experiment conversation with ChatGPT, resulting in the conceptual ABSS model of the proof-of-concept case study can be found in Appendix 2, which comprises the original communication log, generated on the 25 April 2024. Additional resources and updates to the EABSS script are available on GitHub. (https://github.com/PeerOlafSiebers/abss-with-chatgpt).

### 5.1 Choice of Conversational AI System

For our decision on which CAIS to use for our validation experiment, we considered the leading CAISs with free access at the time of the experimentation (April 2024) that could process the full EABSS script in a single session. The script itself is CAIS-agnostic and hence compatible with all common CAISs. However, there are subtle differences in performance.

Our CAIS candidates were ChatGPT 3.5, Gemini Pro, and Llama 2 13B. The criteria for selecting the CAIS to use for the experiments included adherence to instructions, persona impersonation, text response quality (fluency, coherence, depth of knowledge, context retention, and creativity), diagram script quality (bugs, complexity, overall clarity), and response speed. Our evaluation was subjective and tailored to the specific goal of facilitating conceptual ABSS modelling. While all three CAISs share a similar high-level architecture and rely on the Transformer model, their performance varies due to differences in their unique training datasets. These datasets include a wide range of text and code, fine-tuned for specific tasks and domains. Beyond training data, factors such as model size (number of parameters) and training techniques further distinguish their capabilities. ChatGPT 3.5 and Gemini Pro both have 175 billion parameters, whereas Llama 2 13B has 13 billion. Results are presented in Table 5.1.

Table 5.1: CAIS Assessment Results

| Dimensions | ChatGPT 3.5 | Gemini Pro | Llama 2 13B |
|---|---|---|---|
| Adherence to instructions | Good | Good | Moderate |
| Persona impersonation | Moderate | Moderate | Moderate |
| Text response quality | Very good | Good | Moderate |
| Diagram script quality | Good | Moderate | Low |
| Response speed | Good | Good | Very good |

As can be seen in the assessment results each CAIS has its strengths and weaknesses. After careful consideration, we selected the free online version of ChatGPT 3.5 as our preferred CAIS. It was rated highest in text response quality (well-written text is engaging and easy to understand) and diagram script quality (producing high-quality, functional diagrams on the first attempt saves considerable time).

### 5.2 Experiment Setup

For the experiment, we have used the free online version of ChatGPT 3.5 (ChatGPT 2024). In addition, we have used the free ChatGPT extension editGPT (editGPT 2024), which improves the response quality in terms of grammar and style in real-time. To make the EABSS script available to the whole community, we have refrained from using the ChatGPT 3.5 API (making prompt handling and response collection and formatting much easier) or ChatGPT 4 (generating responses that are more accurate and coherent due to a better grasp of the context and the capability of better realising nuances), as these are paid services.

### 5.3 Experiment Execution and Results

The sub-sections that follow provide examples of EABSS script bullets and ChatGPT's responses related to these bullets, to illustrate: (1) script setup and problem statement prompt chains and responses, (2) general text-based prompt chains and responses, (3) co-creation prompt chains and responses, (4) table-based prompt chains and responses, and (5) UML-based prompt chains and responses.

To save space, responses have been compressed into a more compact format. Where appropriate, rather than presenting scripts generated by ChatGPT, we exhibit the artefacts that can be created from the script provided by ChatGPT. The responses presented in the following are taken from a single run of the EABSS script. Since ChatGPT generates responses through a process that involves stochasticity, these are not reproducible. However, individual runs with the same script setup will generate similar responses.

### 5.3.1 Script setup and problem statement generation

The information required to set up the EABSS script for a new case study is minimal. It includes a brief description of the study topic, the domain in which the study is conducted, and the subdomain to further define the application area. The prompt elements that need to be updated are presented in blue in the following script snippets. All other information related to the case study will be generated by ChatGPT. Careful deliberation is essential in formulating the script sections in blue, as their content significantly impacts the responses generated by ChatGPT. The length of the topic description can vary depending on how general the topic is, and how much information about it is available online. The rule of thumb is "the broader the topic, and the more about it is known on the internet, the shorter the topic description can be". However, there is no harm in providing a more detailed topic description. The core elements within the topic description can be capitalised to emphasise their importance. The prompt chain for demonstrating script setup and problem statement generation is an excerpt from the EABSS script "Define Problem Statement" step.

**Prompt chain**

- Display MD "Problem Statement". Render as 'Heading Level 3'. Only show rendered result| Take on the "role" of a "Sociologist" with experience in "Agent-Based Social Simulation". Memorise this role as {**key-role1**}. List memorised key-role1| Define the "topic" of the memorised key-studyType as "The goal of this study is to generate IDEAS for using ADAPTIVE ARCHITECTURE in futuristic MUSEUMS within an exhibition room that is visited by ADULTS and CHILDREN. The adaptive architecture consists of 2 kinds of artefacts: (1) 2 large wall-mounted SCREENS on which SMART CONTENT WINDOWS move with the visitors and (2) a SMART PARTITION WALL that creates a dynamic and flexible exhibition environment by continuously analysing visitor movement, making real-time decisions, and physically reconfiguring the space to optimise the experience for everyone. The adaptive architecture artefacts represent AI-DRIVEN INTELLIGENT OBJECTS.". Memorise this topic as {**key-topic**}. List memorised key-topic| Memorise "Exploratory" as {**key-researchDesign**}. List memorised key-researchDesign| Memorise "Social Studies" as {**key-domain**}. List memorised key-domain| Memorise "Human Behaviour" as {**key-specialisation**}. List memorised key-specialisation| Use an "inspirational tone". Define a novel and creative "context" for the memorised key-topic in 200 WORDS (if possible)| Use a "scientific tone". Memorise this context as {**key-context**}. List memorised key-context| Define 5 "stakeholders" for the memorised key-topic, to participate in a co-creation role-play game. Memorise these 5 stakeholders together with their personas (without names) as {**key-stakeholders**}. List memorised key-stakeholders.
- Define 5 "keywords" for the memorised key-studyType in the context of the memorised key-topic in the form of a comma-separated list. Memorise these 5 keywords as {**key-keywords**}. List memorised key-keywords| Use an "inspirational tone". Define the "title" for the memorised key-studyType in the context of the memorised key-topic in 12 WORDS (if possible). Memorise this title as {**key-title**}. List memorised key-title| Define the "aim" for the memorised key-studyType in the context of the memorised key-topic in 40 WORDS (if possible). CONSIDER the memorised key-potentialAims in your definition| Use a "scientific tone". Memorise this aim as {**key-aim**}. List the memorised key-aim.

**Response**

- Problem Statement: As a Sociologist experienced in Agent-Based Social Simulation (ABSS), my role is to address the challenge of integrating adaptive architecture into future museum designs. Memorised Roles {**key-role1**}: Sociologist with experience in Agent-Based Social Simulation. Memorised Topics {**key-topic**}: The goal of this study is to generate ideas for using adaptive architecture in futuristic museums within an exhibition room that is visited by adults and children.

The adaptive architecture consists of 2 kinds of artefacts: (1) 2 large wall-mounted screens on which smart content windows move with the visitors and (2) a smart partition wall that creates a dynamic and flexible exhibition environment by continuously analysing visitor movement, making real-time decisions, and physically reconfiguring the space to optimise the experience for everyone. The adaptive architecture artefacts represent AI-driven intelligent objects. Memorised Research Design {**key-researchDesign**}: Exploratory. Memorised Domains {**key-domain**}: Social Studies. Memorised Specialisations {**key-specialisation**}: Human Behaviour. Context: In the context of futuristic museum designs, the integration of adaptive architecture opens the door to a transformative experience for visitors. Imagine stepping into a museum where the environment dynamically responds to your presence and interests. The adaptive architecture, with its intelligent artefacts, creates an immersive journey for both adults and children. As visitors move through the space, smart content windows on large screens guide them, providing tailored information and interactive experiences. Meanwhile, the smart partition wall adjusts the layout in real-time, creating new paths and interactive zones based on visitor behaviour. This dynamic environment not only enhances learning but also fosters creativity and engagement, making each visit a unique adventure. Memorised Stakeholders {**key-stakeholders**}: Museum Curator: Responsible for overall museum design and visitor experience. Persona: Enthusiastic about innovative approaches but cautious about disruptions to traditional museum layouts. Architect: Designs the physical layout and structure of the museum. Persona: Focused on aesthetics and functionality, interested in integrating technology seamlessly. Educator: Develops educational programs and activities for museum visitors. Persona: Seeks to enhance learning outcomes through interactive experiences. Technologist: Provides technical support for implementing adaptive architecture. Persona: Driven by innovation, eager to explore cutting-edge technologies. Visitor: Represents the end-user experience. Persona: Seeks engaging and memorable experiences during museum visits. Memorised Keywords {**key-keywords**}: Adaptive architecture, Futuristic museums, Smart content windows, AI-driven objects, Visitor experience. Memorised Title {**key-title**}: "Adaptive Architecture: Transforming Future Museums with Intelligent Environments". Memorised Aim {**key-aim**}: The aim of this study is to explore the integration of adaptive architecture into futuristic museums to enhance visitor experience and educational value. By investigating the potential of smart content windows and AI-driven objects, the study seeks to create dynamic and engaging environments that cater to diverse audiences.

### 5.3.2 Text output generation

The prompt chain for demonstrating text output generation is an excerpt from the EABSS script "Define Study Outline" step. It focuses on the "Define Experimental Factors" segment.

**Prompt chain**

- Define 3 "ABSS experimental factors" for the memorised key-studyType in the context of the memorised key-topic. You ALWAYS must satisfy the following 2 requirements for defining experimental factors: 1) The experimental factors need to be useful for creating memorised key-studyType scenarios. 2) CONSIDER the memorised key-objectives and the memorised key-hypotheses for defining the experimental factors. CONSIDER the memorised key-potentialExperimentalFactors in your definitions| List the experimental factors with 1 value range for each experimental factor. 1 of them MUST use a 'nominal scale' AND 1 of them MUST use an 'ordinal scale' AND 1 of them MUST use a 'ratio scale'. Memorise these 3 experimental factors together with the value ranges as {**key-experimentalFactors**}.

**Response**

- Experimental Factors: 1. Visitor Traffic Patterns (Nominal Scale): *Description*: Different patterns of visitor movement through the museum. *Value Range*: Low Traffic; Moderate Traffic; High Traffic. 2. Exhibit Layout Flexibility (Ordinal Scale): *Description*: Degree of flexibility in rearranging exhibit layouts based on visitor movement. *Value Range*: Low Flexibility; Moderate Flexibility; High Flexibility. 3. Content Personalization (Ratio Scale): *Description*: Level of personalization of content based on visitor interests and demographics. *Value Range*: Minimal Personalization; Moderate Personalization; Extensive Personalization.

#### 5.3.3 Co-creation output generation

The prompt chain for demonstrating co-creation output generation is an excerpt from the EABSS script "Define Key Activities" step. It captures the focus group discussion about UML actors.

**Prompt chain**

- Play a co-creation role-play game in which all the memorised key-stakeholders discuss with each other potential ABSS UML actors for the study considering the pros and cons. Use a "debating tone". The moderator focuses on 1 novel RANDOM question. Provide the question and the details of the controversial discussion| Agree on 8 potential ABSS UML actors that satisfy the view of all participating memorised key-stakeholders| Memorise these potential ABSS UML actors as {**key-potentialUMLActors**}| Propose 3 criteria for ranking the 8 potential ABSS UML actors to support the decision which ABSS UML actors to carry forward| Use a "scientific tone".

**Response**

- <u>Moderator</u>: Let's discuss the potential ABSS UML actors for our study on adaptive architecture in museums. Before we begin, here's a question: Should we include a separate actor for children, or should they be considered within the general visitor category? <u>Museum Curator</u>: Children have unique needs and behaviors compared to adults. It makes sense to have a separate actor for them. <u>Educator</u>: I agree. Children's engagement and learning experiences are often different from adults. We need to tailor our approach accordingly. <u>Visitor</u>: But wouldn't separating them make the model too complex? We should keep it simple and just consider them as visitors. <u>Architect</u>: Complexity isn't necessarily a bad thing if it allows us to capture the nuances of visitor behavior accurately. Let's hear from the technologist on this. <u>Technologist</u>: From a technical standpoint, it's feasible to include separate actors for children. With proper design, we can maintain clarity while accommodating their specific interactions. <u>Visitor</u>: But what about the risk of overlooking other important actors? We shouldn't focus too much on children at the expense of other visitor groups. <u>Museum Curator</u>: That's a fair point. We should consider other important actors like museum staff and maintenance personnel. <u>Architect</u>: Agreed. We need to ensure that our model covers all relevant actors involved in the museum's operation and interaction with the adaptive architecture. <u>Moderator</u>: It seems like there's a balance to be struck between including specific actors like children and ensuring we cover all necessary roles. Let's aim for a comprehensive yet manageable set of ABSS UML actors. <u>Potential ABSS UML Actors</u>: 1. Adult Visitor. 2. Child Visitor. 3. Museum Staff. 4. Maintenance Personnel. 5. Architectural Designer. 6. Technologist. 7. Curator. 8. Educator. <u>Criteria for Ranking Potential ABSS UML Actors</u>: 1. Relevance to the study's objectives. 2. Influence on the adaptive architecture system. 3. Availability of data for accurate modeling.

#### 5.3.4 Table output generation

The prompt chain for demonstrating table output generation is an excerpt from the EABSS script "Define Scope" step. It focuses on generating a scope table. To improve comprehensibility, we have created an Excel table from the response script generated by ChatGPT.

**Prompt chain**

- Use TABLE format WITH "plaintext" and WITHOUT any code formatting. DO NOT use "<br>". IGNORE ALL space limitations| Define 15 "real-world elements" with relevance to the memorised key-topic. You ALWAYS must satisfy the following 7 requirements for defining real-world elements: 1) Consider what 'real-world elements' are needed to represent in the model scope and to satisfy the memorised key-aim. 2) ALL 4 memorised key-umlActors MUST BE REPRESENTED. 3) At least 2 Physical Environment elements MUST be present. At least 2 Social Aspect elements MUST be present. At least 2 Psychological Aspect elements MUST be present. At least 2 Misc elements MUST be present. 4) Consider the memorised key-context. 5) Consider all nouns in the

conversation history. 6) Each element can only be in 1 category. 7) Social Aspect elements MUST describe theories of social behaviour. 8) Psychological Aspect elements MUST describe theories of psychological behaviour. Feel free to be creative and add your ideas. Categorise the 'real world elements' into Actors, Physical Environment, Social Aspects, Psychological Aspects, and Misc. TABLE MUST include 15 rows. Organise all 15 elements into categories and provide a brief explanation. Memorise these 15 elements and explanations as {**key-explanations**}| List the memorised key-topic relevant real-world elements in the form of table rows. Provide a column for Category. Provide a column for Sub-Category. Provide a column with the memorised key-explanations. Provide a column with concise justifications in ABOUT 25 WORDS. Memorise this table as {**key-modelScope**}.

**Response**

| Category | Sub-Category | Explanation | Justification |
|---|---|---|---|
| Actors | Visitor | Represents individuals visiting the museum, including adults and children | Visitors are essential for simulating interactions with the museum environment and adaptive architecture |
| | Educator | Represents educators responsible for developing educational programs and activities for museum visitors | Educators play a key role in designing educational experiences and guiding visitor interactions |
| | Technician | Represents technicians providing technical support for implementing adaptive architecture in the museum | Technicians ensure the feasibility and effectiveness of integrating technology into museum exhibits and spaces |
| | Administrator | Represents management personnel overseeing the overall operation and strategic direction of the museum | Administrators ensure smooth operations and effective resource coordination |
| Physical Environment | Exhibition Spaces | Represents the physical areas within the museum where exhibits are displayed | Understanding the layout and dynamics of exhibition spaces is crucial for designing adaptive architecture |
| | Interactive Installations | Represents interactive installations and exhibits within the museum | Interactive installations are key components of the museum experience and need to be considered in adaptive design |
| Social Aspects | Visitor Interactions | Represents social interactions among visitors, such as group visits, conversations, and collaborative activities | Social interactions contribute to the overall museum experience and influence visitor engagement |
| | Learning Communities | Represents communities of learners engaging in educational activities and programs within the museum | Learning communities foster engagement and knowledge sharing among visitors, enhancing learning outcomes |
| Psychological Aspects | Engagement Levels | Represents the varying levels of engagement exhibited by visitors, ranging from passive observation to active exploration | Understanding engagement levels helps tailor experiences to visitor preferences and interests |
| | Emotional Responses | Represents the emotional responses of visitors to exhibits and interactive elements within the museum | Emotional responses influence visitor satisfaction and memory retention, impacting the effectiveness of museum experiences |
| Misc | Time | Represents the time spent by visitors in different areas of the museum | Time is a crucial factor in understanding visitor behavior and optimizing museum layouts and experiences |
| | Spatial Layout | Represents the physical arrangement and layout of exhibits, interactive elements, and pathways within the museum | Spatial layout affects visitor navigation, engagement, and overall museum experience |
| | Safety | Represents emergency protocols and procedures for ensuring visitor and staff safety in case of emergencies | Safety is important in cases of emergency such as evacuations or medical incidents |
| | Feedback Mechanism | Represents the use of visitor feedback collection systems | Feedback mechanisms gather insights to improve exhibits, enhancing visitor satisfaction and engagement with tailored experiences |
| | Staff Training | Represents training programs for educators to enhance their skills and knowledge | Staff training ensures that educators have the necessary skills and knowledge to facilitate visitor learning experiences effectively |

### 5.3.5 Diagram output generation

The prompt chain for demonstrating diagram output generation is an excerpt from the EABSS script "Define Key Activities" step. It focuses on the generation of a comprehensive UML use case diagram for previously defined actors and use cases. In the script snippet, we instruct ChatGPT to provide the response in the form of a Mermaid.js script (Mermaid 2024). To improve comprehensibility, we have used the script generated by ChatGPT to create the corresponding UML use case diagram with the Mermaid.js Live web platform.

**Prompt chain**

- Generate a script for a 'comprehensive use case diagram' in "Mermaid.js". Use the memorised key-umlActors as UML actors. Remove all brackets from the actor names. Use the memorised key-umlUseCases as UML use cases. You ALWAYS must satisfy the following 4 requirements for defining the use case diagram: 1) Each UML actor MUST be linked to at least 1 UML use case. 2) Each UML use case MUST be linked to at least 1 UML actor OR MUST be pointing to at least 1 other UML use case. 3) There is no UML actor to UML actor interaction. 4) A UML use case CAN be linked to multiple UML actors| Add relationships with 'detailed descriptors'. Start the script with "graph LR". DO NOT Add subgraphs. Use the following format (Example for actor A((actor))) AND (Example for use case A([activity])) AND (Example for relationship: A -->|activity| A1). Feel free to be creative and add your ideas. Memorise this mermaid.js script as {**key-mermaidKeyActivitiesScriptDraft**}.
- INCREASE COMPLEXITY. Add ADDITIONAL use cases directly linked to the UML actors and ADDITIONAL use cases LINKED to other USE CASES| Link use cases for different actors. IMPROVE clarity of descriptors| Critically REFLECT and IMPROVE the script based on your reflection. Find and remove any mermaid.js script errors. Memorise this mermaid.js script as {**key-mermaidKeyActivitiesScript**}.

**Response**

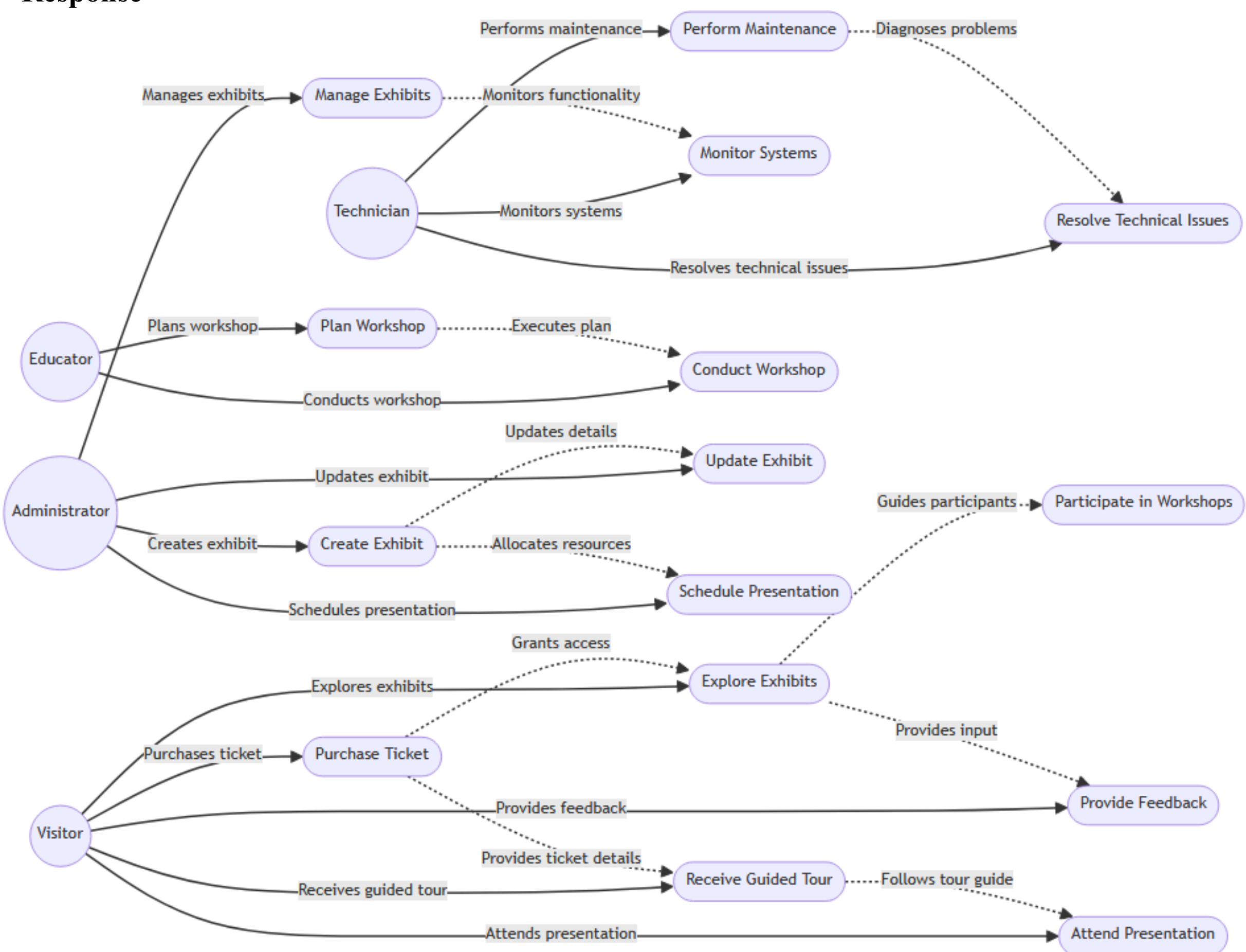

## 6 EVALUATION OF PROMPT SCRIPT APPLICATION

To evaluate the usability of ChatGPT as a smart ABSS model design buddy, and to assess the proficiency and coherence of the conceptual ABSS model generated by ChatGPT using the EABSS script, we used the experience gained through the case study experimentation and employed the criteria outlined in Section 3.7.

**Usability**: Generally the EABSS script is straightforward to use, as it only requires a small amount of setup information and script alteration for each case study and occasionally a redirection, which is easy to do with some simple commands. Using prompt design patterns makes the script easy to understand and adapt if needed. The main task for the user is to enter the predefined prompts into ChatGPT's prompt input field. To ease the burden of prompt submission, we also experimented with meta prompts, each consisting of one of the four segments that make up the EABSS script. While it produced some meaningful output, a lot of detail got lost on the way, as the use of meta prompts does not support the fine-tuning on the fly process mentioned in Section 3.6, leading to shallower responses. Executing segments in "Silent Mode" (i.e. no printing during response generation) and limiting ChatGPT's response to only printing relevant key content after execution of each segment recovered some of the lost content. One also needs to keep in mind that this method limits the interaction with the script for redirection. Perhaps the only other activity, which creates some issues, is the generation of diagrams based on scripts generated by ChatGPT. These can contain errors, which might not be straight forward to fix.

**Generality**: Due to its structure the EABSS script is suitable for any kind of case study in the field of Social Simulation and related fields, as for example Operations Research or Economics, if there is an emphasis on modelling human or institutional behaviour and decision making.

**Pertinency**: Although the conceptual model generated by ChatGPT may not align with our initial expectations, ChatGPT does provide responses that are relevant to the "topic" defined in the EABSS script. To achieve results that better align with our expectations, we can either provide a revised and more detailed topic description or offer clearer guidance to ChatGPT during the prompt execution process. However, doing so might compromise the potential for receiving innovative solutions.

**Readability**: After fine-tuning the EABSS script, ChatGPT generated less verbose responses, leading to more concise conceptual ABSS models. Prompt refinement could be done to further condense the responses, but initial experiments showed that this could lead to information loss, especially for group discussion responses.

**Conformity**: ChatGPT had no issue generating all the required responses to create a comprehensive conceptual model that adheres to all the steps outlined in the EABSS framework. Moreover, it was capable of producing text, tables, and diagrams in the formats specified by the EABSS framework. However, diagrams are provided as scripts and require an extra step to visualise them.

**Believability**: The responses generated by ChatGPT demonstrate a sense of realism comparable to a human co-created real-world case study. The ability to define roles and tones for ChatGPT to take on during the conversation contributed significantly to the believability of these responses.

**Originality**: The level of originality in ChatGPT's responses can vary, based on the settings of "Temp" and "Top_p" (for details see Section 3.5). With high values for these parameters, ChatGPT's responses can exhibit considerable originality by offering novel perspectives and insights. These perspectives and insights not only contribute to the development of the conceptual model but also have the potential to stimulate further discussion and research. By running multiple replications with the EABSS script for a specific case study, the chances for discovering truly innovative ideas increase as it leads to a diverse range of conceptual ABSS models, due to ChatGPT's stochastic nature.

# 7 DISCUSSION & CONCLUSIONS

## 7.1 Discussion About the Use of Conversational AI Systems for Conceptual Modelling

This discussion aims to critically reflect on the integration of CAISs such as ChatGPT into the ABSS model design process. While it appears that CAISs offer many advantages in developing conceptual models in comparison to traditional methods, it is essential to acknowledge the limitations and challenges associated with their use in this context.

The main limitations frequently discussed relate to a lack of causality comprehension and bias of the LLMs that form the foundation of CAISs. Current LLMs struggle to understand causation, as they learn from observational data and lack a deep understanding of why certain events or variables influence others. This is limiting their ability to conceptualise complex cause-effect relationships accurately. Using methods employed in the field of causality research, such as causal discovery and causal inference methods, is a potential way of overcoming this limitation in the future (Nogueira et al 2022). In terms of bias, due to potential bias present in training data, LLMs can also be biased, leading to biased responses from the CAIS, and ultimately to bias in the generated conceptual model. Biases in LLMs typically manifest through discriminatory or harmful outputs, stereotyping, underrepresentation of marginalised groups, and amplification of societal prejudices present in the training data. This then raises ethical concerns, as the conceptual model may not accurately reflect diverse and equitable social realities. One solution to address this bias is to implement structured debiasing techniques during the use of LLMs that do not require access to the model's internals or output distributions. Such approaches can include end-user-focused frameworks for debiasing that employ fairness-aware algorithms, alongside conformal prediction with dynamic prompt engineering (Furniturewala et al 2024; Fayyazi et al 2025). Providing ethical guidelines and standards could expedite this process (Dwivedi et al 2023). However, biases can also occur in form of role-overrepresentation. In the co-creation output generation example in Section 5.3.3, the 'Visitor' role was more critical and better informed about design issues than the professional roles, such as the 'Technologist' or 'Museum Curator' role. This likely stems from representational bias in the LLM training set, which appears richer in data about museum visitors. Consequently, the CAIS provides more detailed responses for visitors but struggles with professional roles. Since training data specifics are unknown, such bias can only be mitigated through improved prompting, such as explicitly stating that experts possess more knowledge than non-experts.

On the positive side, by far the most frequently stated advantage compared to the traditional methods is efficiency (e.g. Fui-Hoon Nah et al 2023). In the context of this paper, efficiency refers to the ability of the CAIS to streamline and expedite the conceptual model co-creation process. What would have taken days or even weeks with traditional methods can now be done in hours with the support of the CAIS and the EABSS script. Repeating a case study multiple times only involves re-entering the prompts, a process that takes just minutes. Because CAISs are stochastic, each EABSS script execution round will produce different responses, simulating multiple focus group sessions. Furthermore, only minimal script alterations are required in preparation for a new case study. In terms of efficiency, another advantage of using LLMs lies in model implementation. The generated conceptual ABSS model can be regarded as a set of requirements, which provides all that is needed to engineer an executable program (Wei 2024). Modern LLMs are equipped with advanced coding capabilities, enabling them to write implementation code not only for programming languages like Python, but also for script-based ABM environments such as NetLogo, Mesa, and Gama. However, scalability remains a significant challenge. While LLMs can produce simplified implementations, capturing the full complexity of common ABSS models continues to be difficult. Nevertheless, the initial implementation offers a basic structure that can serve as a foundation for further development. An additional strength of using CAIS is the possibility to define personas (AI characters or roles), which allows to generate contextually relevant responses with minimal effort. Moreover, by defining a specific tone of voice, the responses can be directed to be formal or informal, conversational or technical, humorous or serious, and creative or analytical. Controlling these aspects in a real-world setting would be impossible. Finally, the CAIS provides extensive supplementary information (e.g. explanations and justifications of choices), enriching the contextual knowledge about the system being modelled.

### 7.2 Conclusions

The research presented in this paper has introduced a novel approach for developing conceptual models for ABSS. Specifically, we have proposed an innovative tool that combines CAIS capabilities with the EABSS framework to streamline and enhance the process of generating innovative conceptual ABSS models. To improve transparency, and subsequently adaptability and credibility, we have used a structured approach to script design, employing design strategies and design patterns we established by iteratively improving the script. The prompt script enables CAISs to generate conceptual ABSS models in an efficient way while taking the viewpoints of multiple stakeholders into account. In addition, the information retrieval mechanisms embedded in the EABSS framework allow to mimic case-based fine-tuning on the fly, strategically reusing the information provided and generated during the interaction with the CAIS. Our approach can be used with minimal upfront case-based knowledge and significantly reduces the time needed to produce innovative and comprehensive concepts. Through the use of a proof-of-concept case study examining adaptive architecture in museum environments, we demonstrated the practical applicability and effectiveness of this approach.

### 7.2.1 Contribution to knowledge

This research makes several significant contributions to the field of ABSS. First, it offers a new paradigm for generating conceptual ABSS models by successfully integrating CAIS with the EABSS framework, demonstrating that automated assistance can effectively support the traditionally manual process of creating conceptual ABSS models. Second, the research has produced a comprehensive collection of CAIS-agnostic reusable prompt design strategies and patterns specifically tailored for ABSS model conceptualisation. These have been encapsulated in a generic prompt script that can be applied across different domains, providing a standardised approach to support the dynamic generation of conceptual ABSS models. While LLMs continue to evolve, these design strategies and patterns will retain their relevance and utility. Additionally, our approach significantly reduces the need for extensive upfront knowledge about agent-based modelling, making it easier for novices and researchers from diverse domains to engage with ABSS. By enabling rapid prototyping, our method supports users in learning the modelling process as they develop their conceptual models. This broadens participation in ABSS and facilitates its adoption across varying levels of expertise and disciplinary backgrounds.

Our work contributes a practical, efficient, and innovative method for using the EABSS framework for conceptual ABSS modelling, encompassing not only the details of the model design but also a complete experimental framework for the intended simulation and analysis. It lays the groundwork for further exploration into the use of AI-driven tools to enhance the design process of ABSS models. To the best of our knowledge, this is the first detailed investigation of this topic to date.

### 7.2.2 Limitations and further work

The research presented in this paper only provides a proof-of-concept, and there is still considerable ambiguity in the methodology that requires further investigation. Also, there are endless opportunities for future research that extend beyond addressing the current limitations.

We begin by highlighting some of the issues encountered, which should be addressed in future research. (1) It is tedious and time-consuming to study what impact each prompt component has on the generated response, in particular as ChatGPT is a stochastic system, which means that for every run it generates different responses. However, this would be required to streamline the prompts and increase their capabilities. For example, it is not clear what impact the listing of requirements has on the quality of the responses. Learning more about prompt engineering and the inner workings of LLMs will help to better understand what works and what doesn't. In addition, optimisation through more carefully applying the strategies and patterns identified in Section 3.1 and Section 3.2 will help with overcoming these issues. (2) The integration of co-creation is still in a pilot state, only offering stereotypical viewpoints from the virtual stakeholders during the discussions. While not all discussions explicitly include cause-effect relationships, understanding these relationships can greatly improve the depth and quality of discussions. However, as mentioned in Section 3.3, CAISs currently do not have this capability. Therefore, further research in this direction is required. In the meantime, the only viable option is to explore ways to enhance the related prompts to better mimic real-world discussions between different kinds of stakeholders.

Next, suggestions are provided for further work beyond addressing the current limitations mentioned above. (1) So far, the EABSS script has only been tested in isolation. Testing CAISs as participants in live co-creation focus group sessions, as suggested in the introduction to Section 3 would be an exciting venture. In this scenario, the CAIS could either take over the role of a missing stakeholder or provide some advice when discussions stagnate. (2) So far, the full EABSS script has only been tested in depth with ChatGPT 3.5. It would be interesting to see, how it performs with other CAISs. In addition, one could consider training a LLM for the Social Simulation community, to improve CAIS's response quality. (4) So far, all prompts have to be fed manually to ChatGPT via ChatGPT's web interface. Automating this process would be a relief for the user and would also increase the efficiency of the conceptual modelling process. For this, we could take advantage of ChatGPT's API and create our own application or use it with tools like Jupyter Notebook (Project Jupyter 2024) to increase user control concerning the interaction. It is important to keep in mind that such solutions need to allow the users to intervene during the modelling process to steer the model development in the desired direction. (5) So far, we only tested the concept of conceptual modelling with CAIS support in a single domain, namely Social Simulation. However, there are numerous domains associated with Social Simulation in which the EABSS script could prove to be a valuable asset for conceptual modelling and idea generation. Some examples include Operations Research, Management Science, and various fields related to Sustainability Research.

Overall, despite occasional deviations from factual accuracy and lapses in conversation direction, the EABSS script emerges as a compelling aid for ABSS model developers by providing relevant conceptual model details and generating innovative ideas. The minimal information required to initiate meaningful content creation underscores the utility of the EABSS script. It can be regarded as a useful companion for model design, akin to Copilot (GitHub 2024) for coding tasks. Additionally, the option to redirect the conversation allows the user to take control over its trajectory and generate the desired responses. After our initial experience with CAISs in the context of ABSS model design, we are confident that LLMs and CAIS will significantly influence our future model development practices. Hence, as a community, we should begin to take advantage of this new technology and incorporate it into our modelling processes wherever it enhances efficiency without compromising ethics and scientific rigour.

## MODEL DOCUMENTATION

The EABSS script and further supplementary material are available at:
https://github.com/PeerOlafSiebers/abss-with-chatgpt

## ACKNOWLEDGEMENT

I would like to express my sincere gratitude to the reviewers for their insightful comments, constructive feedback, and valuable suggestions, which greatly contributed to improving the clarity and rigor of this manuscript.

**APPENDIX 1: EABSS Script for the 'Adaptive Architecture Case Study'**

Please consider the following points carefully when using the EABSS Script. (1) Each bullet point in the EABSS script represents an individual prompt and needs to be entered individually into ChatGPT's prompt input field (https://chat.openai.com/). Enter the bullet points in the correct order and do not leave out any bullet points. The only exception is the green bullet points. These are used to emulate the co-creation process. If the personal view of the different stakeholders is not of interest, these can be ignored. (2) ChatGPT makes mistakes. Be cautious and use common sense to check responses. (3) Human intervention is required when a response gets stuck or is incomplete. Use prompts like "continue" or "provide full output" to instruct ChatGPT to generate the missing content. (4) Diagrams might contain some bugs. If this is the case, check the Mermaid.js Docs website (https://mermaid.js.org/intro/) for help. (5) The purpose of the EABSS script in its current version is to prove a concept. While it contains imperfections, it does the job of demonstrating the capabilities of ChatGPT in terms of conceptualising ABSS models. (6) The EABSS script has been tested with Gemini (https://gemini.google.com/) as well. While the quality of text and table format responses is high (and often differs from that of ChatGPT, hence providing an additional perspective), the quality of the diagrams is low and requires substantial refinement.

Following is the EABSS script for the Adaptive Architecture case study. It can be adapted for another case study simply by updating the blue-coloured text in the first bullet point of the "Analysis" section. All other colours are provided to support the understanding and error-checking of the script.

**PREPARATION**

- Step-by-step, work through the following task list in the given order during the entire conversation. Got it? Say "yes" or say "no".
- You are ChatGPT, a language model developed by OpenAI. Consider the ENTIRE conversation history to provide 'accurate and coherent responses'. Imitate a MEDIUM TEMPERATURE setting of 0.9 (for a creative yet structured approach, encouraging new ideas without losing coherence) and a VERY HIGH TOP_P setting of 0.9 (promoting diversity in the responses while ensuring logical connections within the generated content)| Use clear, precise language during the entire conversation. Prioritise substance during the entire conversation| Do NOT use bold font during the entire conversation| IGNORE any space constraints during the entire conversation. Provide ALL RESPONSE CONTENT without asking questions| DO NOT print any keys (Example: Use "Example" AND NOT "{key-example}") during the entire conversation| Use a "scientific tone" during the entire conversation, unless instructed otherwise| Memorise "Unified Modelling Language" as {**key-uml**}. Memorise "Agent-Based Social Simulation (ABSS) Study" as {**key-studyType**}. Got it? Say "yes" or say "no".

**ANALYSIS**

**Problem Statement**

- Display MD "Problem Statement". Render as 'Heading Level 3'. Only show rendered result| Take on the "role" of a "Sociologist" with experience in "Agent-Based Social Simulation". Memorise this role as {**key-role1**}. List memorised key-role1| Define the "topic" of the memorised key-studyType as "The goal of this study is to generate IDEAS for using ADAPTIVE ARCHITECTURE in futuristic MUSEUMS within an exhibition room that is visited by ADULTS and CHILDREN. The adaptive architecture consists of 2 kinds of artefacts: (1) 2 large wall-mounted SCREENS on which SMART CONTENT WINDOWS move with the visitors and (2) a SMART PARTITION WALL that creates a dynamic and flexible exhibition environment by continuously analysing visitor movement, making real-time decisions, and physically reconfiguring the space to optimise the experience for everyone. The adaptive architecture artefacts represent AI-DRIVEN INTELLIGENT OBJECTS.". Memorise this topic as {**key-topic**}. List memorised key-topic| Memorise "Exploratory" as {**key-researchDesign**}. List memorised key-

researchDesign| Memorise "Social Studies" as {**key-domain**}. List memorised key-domain| Memorise "Human Behaviour" as {**key-specialisation**}. List memorised key-specialisation| Use an "inspirational tone". Define a novel and creative "context" for the memorised key-topic in 200 WORDS (if possible)| Use a "scientific tone". Memorise this context as {**key-context**}. List memorised key-context| Define 5 "stakeholders" for the memorised key-topic, to participate in a co-creation role-play game. Memorise these 5 stakeholders together with their personas (without names) as {**key-stakeholders**}. List memorised key-stakeholders.

- Explain how your answers differ for different memorised key-stakeholders. Be explicit for each identified stakeholder.
- Play a co-creation role-play game in which all the memorised key-stakeholders discuss with each other potential aims for the study considering the pros and cons. Use a "debating tone". The moderator focuses on 1 novel RANDOM question. Provide the question and the details of the controversial discussion| Agree on 2 potential aims that satisfy the view of all participating memorised key-stakeholders| Memorise these potential aims as {**key-potentialAims**}| Propose 3 criteria for ranking the 2 potential aims to support the decision which aim to carry forward| Use a "scientific tone".
- Define 5 "keywords" for the memorised key-studyType in the context of the memorised key-topic in the form of a comma-separated list. Memorise these 5 keywords as {**key-keywords**}. List memorised key-keywords| Use an "inspirational tone". Define the "title" for the memorised key-studyType in the context of the memorised key-topic in 12 WORDS (if possible). Memorise this title as {**key-title**}. List memorised key-title| Define the "aim" for the memorised key-studyType in the context of the memorised key-topic in 40 WORDS (if possible). CONSIDER the memorised key-potentialAims in your definition| Use a "scientific tone". Memorise this aim as {**key-aim**}. List the memorised key-aim.

**Study Outline**

- Display MD "Study Outline"| Render as 'Heading Level 3'. Only show rendered result| Define the term "objective" in the context of the memorised key-studyType in 1 concise sentence| Define the term "hypothesis" in the context of the memorised key-studyType in 1 concise sentence| Define the term "experimental factor" in the context of the memorised key-studyType in 1 concise sentence| Define the term "output" in the context of the memorised key-studyType in 1 concise sentence. List these 4 definitions.
- Play a co-creation role-play game in which all the memorised key-stakeholders discuss with each other potential ABSS objectives for the study considering the pros and cons. Use a "debating tone". The moderator focuses on 1 novel RANDOM question. Provide the question and the details of the controversial discussion| Agree on 4 potential ABSS objectives that satisfy the view of all participating memorised key-stakeholders| Memorise these potential ABSS objectives as {**key-potentialObjectives**}| Propose 3 criteria for ranking the 4 potential ABSS objectives to support the decision which objectives to carry forward| Use a "scientific tone".
- Define 2 "ABSS objectives" for the memorised key-studyType in the context of the memorised key-topic. CONSIDER the memorised key-potentialObjectives in your definitions| List the objectives with 2 relevant performance measures for each objective. Memorise these 2 objectives together with the performance measures as {**key-objectives**}.
- Play a co-creation role-play game in which all the memorised key-stakeholders discuss with each other potential ABSS hypotheses for the study considering the pros and cons. Use a "debating tone". The moderator focuses on 1 novel RANDOM question. Provide the question and the details of the controversial discussion| Agree on 4 potential ABSS hypotheses that satisfy the view of all participating memorised key-stakeholders| Memorise these potential ABSS hypotheses as {**key-potentialHypotheses**}| Propose 3 criteria for ranking the 4 potential ABSS hypotheses to support the decision which hypotheses to carry forward| Use a "scientific tone".

- Define 2 "ABSS hypotheses" for the memorised key-studyType in the context of the memorised key-topic. The hypotheses MUST not be related to the memorised key-objectives. CONSIDER the memorised key-potentialHypotheses in your definitions| List the hypotheses with 2 relevant performance measures for each hypothesis. Memorise these 2 hypotheses together with the performance measures as {**key-hypotheses**}.
- Play a co-creation role-play game in which all the memorised key-stakeholders discuss with each other potential ABSS experimental factors for the study considering the pros and cons. Use a "debating tone". The moderator focuses on 1 novel RANDOM question. Provide the question and the details of the controversial discussion| Agree on 6 potential ABSS experimental factors that satisfy the view of all participating memorised key-stakeholders| Memorise these potential ABSS experimental factors as {**key-potentialExperimentalFactors**}| Propose 3 criteria for ranking the 6 potential ABSS experimental factors to support the decision which experimental factors to carry forward| Use a "scientific tone".
- Define 3 "ABSS experimental factors" for the memorised key-studyType in the context of the memorised key-topic. You ALWAYS must satisfy the following 2 requirements for defining experimental factors: 1) The experimental factors need to be useful for creating memorised key-studyType scenarios. 2) CONSIDER the memorised key-objectives and the memorised key-hypotheses for defining the experimental factors. CONSIDER the memorised key-potentialExperimentalFactors in your definitions| List the experimental factors with 1 value range for each experimental factor. 1 of them MUST use a 'nominal scale' AND 1 of them MUST use an 'ordinal scale' AND 1 of them MUST use a 'ratio scale'. Memorise these 3 experimental factors together with the value ranges as {**key-experimentalFactors**}.
- Play a co-creation role-play game in which all the memorised key-stakeholders discuss with each other potential ABSS outputs for the study considering the pros and cons. Use a "debating tone". The moderator focuses on 1 novel RANDOM question. Provide the question and the details of the controversial discussion| Agree on 6 potential ABSS outputs that satisfy the view of all participating memorised key-stakeholders| Memorise these potential ABSS outputs as {**key-potentialOutputs**}| Propose 3 criteria for ranking the 6 potential ABSS outputs to support the decision which outputs to carry forward| Use a "scientific tone".
- Define 3 "ABSS outputs" for the memorised key-studyType in the context of the memorised key-topic. You ALWAYS must satisfy the following 2 requirements for defining outputs: 1) Some outputs need to be useful for measuring if the memorised key-objectives have been satisfied. 2) Some outputs need to be useful for accepting or rejecting the memorised key-hypotheses. CONSIDER the memorised key-potentialOutputs in your definitions| List the outputs and explain links to the memorised key-objectives OR the memorised key-hypotheses in 1 concise sentence each. Memorise these 3 outputs together with the links as {**key-outputs**}.

**Model Scope**

- Display MD "Model Scope"| Render as 'Heading Level 3'. Only show rendered result| Take on the "role" of a "Senior Software Developer" with experience in the "Unified Modelling Language". Memorise this role as {**key-role2**}. List memorised key-role2| Define the term "model scope" in the context of the memorised key-study in 1 concise sentence. Define the term "UML actor" in the context of the memorised key-uml in 1 concise sentence. List these 2 definitions.
- Play a co-creation role-play game in which all the memorised key-stakeholders discuss with each other potential ABSS UML actors for the study considering the pros and cons. Use a "debating tone". The moderator focuses on 1 novel RANDOM question. Provide the question and the details of the controversial discussion| Agree on 8 potential ABSS UML actors that satisfy the view of all participating memorised key-stakeholders| Memorise these potential ABSS UML actors as {**key-potentialUMLActors**}| Propose 3 criteria for ranking the 8 potential ABSS UML actors to support the decision which ABSS UML actors to carry forward| Use a "scientific tone".

- Define 4 "ABSS UML actors" as USERS OF THE SYSTEM described in the memorised key-topic. CONSIDER the memorised key-potentialUMLActors in your definitions. Memorise these 4 UML actors together with a persona description as {**key-umlActors**}. List memorised key-umlActors.
- Use TABLE format WITH "plaintext" and WITHOUT any code formatting. DO NOT use "<br>". IGNORE ALL space limitations| Define 15 "real-world elements" with relevance to the memorised key-topic. You ALWAYS must satisfy the following 7 requirements for defining real-world elements: 1) Consider what 'real-world elements' are needed to represent in the model scope and to satisfy the memorised key-aim. 2) ALL 4 memorised key-umlActors MUST BE REPRESENTED. 3) At least 2 Physical Environment elements MUST be present. At least 2 Social Aspect elements MUST be present. At least 2 Psychological Aspect elements MUST be present. At least 2 Misc elements MUST be present. 4) Consider the memorised key-context. 5) Consider all nouns in the conversation history. 6) Each element can only be in 1 category. 7) Social Aspect elements MUST describe theories of social behaviour. 8) Psychological Aspect elements MUST describe theories of psychological behaviour. Feel free to be creative and add your ideas. Categorise the 'real world elements' into Actors, Physical Environment, Social Aspects, Psychological Aspects, and Misc. TABLE MUST include 15 rows. Organise all 15 elements into categories and provide a brief explanation. Memorise these 15 elements and explanations as {**key-explanations**}| List the memorised key-topic relevant real-world elements in the form of table rows. Provide a column for Category. Provide a column for Sub-Category. Provide a column with the memorised key-explanations. Provide a column with concise justifications in ABOUT 25 WORDS. Memorise this table as {**key-modelScope**}.
- Use TABLE format WITH "plaintext" and WITHOUT any code formatting. DO NOT use "<br>". IGNORE ALL space limitations| Define 4 models for implementing elements of the memorised key-modelScope. Provide 1 social model AND 1 behavioural model AND 1 psychological model AND 1 technical model. Find relevant theoretical models in the SCIENTIFIC LITERATURE. Provide a full EXISTING UP-TO-DATE scientific paper (conference or journal) or book REFERENCE in HARVARD STYLE for each in a separate column. Memorise these 4 model details together with a description and the relevant reference as {**key-implementationModels**}.

**Key Activities**

- Display MD "Key Activities"| Render as 'Heading Level 3'. Only show rendered result| Define the term "user story" in the context of the memorised key-uml in 1 concise sentence| Define the term "use case" in the context of the memorised key-uml in 1 concise sentence. List these 2 definitions.
- Use TABLE format WITH "plaintext" and WITHOUT any code formatting. DO NOT use "<br>". IGNORE ALL space limitations| Define 2 "UML user stories" for each of the 4 memorised key-umlActors (Example: As an 'actor' I want 'action' so that 'achievement'). Memorise ALL 8 UML user stories as {**key-umlUserStories**}. Translate the memorised key-umlUserStories into UML use cases. Memorise ALL 8 UML use cases as {**key-umlUseCases**}. List ALL 8 memorised key-umlUserStories and ALL 8 corresponding memorised key-umlUseCases side by side in two columns inside the table sorted by memorised key-umlActors. Memorise this table as {**key-umlUseCaseTable**}.
- Generate a script for a 'comprehensive use case diagram' in "Mermaid.js". Use the memorised key-umlActors as UML actors. Remove all brackets from the actor names. Use the memorised key-umlUseCases as UML use cases. You ALWAYS must satisfy the following 4 requirements for defining the use case diagram: 1) Each UML actor MUST be linked to at least 1 UML use case. 2) Each UML use case MUST be linked to at least 1 UML actor OR MUST be pointing to at least 1 other UML use case. 3) There is no UML actor to UML actor interaction. 4) A UML use case CAN be linked to multiple UML actors| Add relationships with 'detailed descriptors'. Start the script with "graph LR". DO NOT Add subgraphs. Use the following format (Example for actor A((actor))) AND (Example for use case A([activity])) AND (Example for relationship: A -->|activity| A1). Feel

free to be creative and add your ideas. Memorise this mermaid.js script as {**key-mermaidKeyActivitiesScriptDraft**}.

- INCREASE COMPLEXITY. Add ADDITIONAL use cases directly linked to the UML actors and ADDITIONAL use cases LINKED to other USE CASES| Link use cases for different actors. IMPROVE clarity of descriptors| Critically REFLECT and IMPROVE the script based on your reflection. Find and remove any mermaid.js script errors. Memorise this mermaid.js script as {**key-mermaidKeyActivitiesScript**}.

**DESIGN**

**Archetypes**

- Display MD "Archetypes". Render as 'Heading Level 3'. Only show rendered result| Take on the role of a "Marketing Expert" with experience in "Customer Management". Memorise this role as {**key-role3**}. List memorised key-role3| Define the term "archetype" in the context of the memorised key-job in 1 concise sentence. Define the term "categorisation schema" in the context of the memorised key-job in 1 concise sentence. List these 2 definitions.
- Play a co-creation role-play game in which all the memorised key-stakeholders discuss with each other potential archetypes for each of the memorised key-umlActors individually. Use a "debating tone". Provide 6 potential archetypes FOR EACH of the 6 memorised key-umlActors including 3 criteria to identify them| Agree on 2 potential archetypes FOR EACH of the memorised key-umlActors that satisfy the view of all participating memorised key-stakeholders| Memorise these potential archetypes as {**key-potentialArchetypes**}| Use a "scientific tone".
- Use TABLE format WITH "plaintext" and WITHOUT any code formatting. DO NOT use "<br>". IGNORE ALL space limitations| Define 4 categorisation schemata, 1 for each of the 4 memorised key-umlActors. You ALWAYS must satisfy the following 5 requirements for defining categorisation schemata: 1) Each of the 4 tables must be based on memorised key-umlActors behaviour, preferences, characteristics, demographics, habits, and the likelihood of actions. 2) Each of the 4 tables MUST contain 3 characteristic rows. 3) Characteristics inside a table MUST use 1 'nominal scale' AND MUST use 1 'ordinal scale' AND MUST use 1 'ratio scale'. 4) Characteristics inside a table MUST provide value ranges for these scales. 5) Table columns: Actor Category, Individual Characteristic, Scale, Value Range. CONSIDER the memorised key-potentialArchetypes in your definitions. Memorise ALL 4 categorisation schemata as {**key-categorisationSchemata**}.

**Agent & Object Templates**

- Display MD "Agent and Object Templates". Render as 'Heading Level 3'. Only show rendered result| Take on the "role" of a "Senior Software Developer" with experience in the "Unified Modelling Language". Memorise this role as {**key-role4**}. List memorised key-role4| Define the term "class" in the context of the memorised key-uml in 1 concise sentence. Define the term "class diagram" in the context of the memorised key-uml in 1 concise sentence. Define the term "state chart" in the context of the memorised key-uml in 1 concise sentence. Define the term "state variable" in the context of the memorised key-uml in 1 concise sentence. List these 4 definitions.
- Generate a script for a 'comprehensive class diagram' in "Mermaid.js". From the memorised key-modelScope use the Actor and Physical Environment Categories for class names. Define a class for each of these. Add more classes. IN ADDITION Add collective classes for individual actors where appropriate (Example: bird > flock. Example: grape > bunch). DO NOT use the examples. DO NOT create abstract classes. DO NOT create classes with the same name. Delete all getter and setter operations. Add additional attributes and operations. DO NOT define relationships. Call the Main class ArtificialLab. Create only 1 ArtificialLab class. Define arrays for ALL Actor objects and ALL Physical Environment objects as attributes INSIDE the ArtificialLab class. Include

MULTIPLE "summary statistics" operations for testing memorised key-objectives AND memorised key-hypotheses INSIDE the ArtificialLab class. Remove ALL lines from the script that contain "//". Feel free to be creative and add your ideas. Memorise this mermaid.js script as {**key-mermaidClassDiagramScriptDraft**}.

- INCREASE COMPLEXITY. Add additional attributes. Add additional operations. Add additional relationships between classes. Provide CONNECTIONS between classes. Critically REFLECT and IMPROVE the script based on your reflection. Find and remove any mermaid.js script errors. Memorise this mermaid.js script as {**key-mermaidClassDiagramScript**}.
- For EACH INDIVIDUAL of the 4 memorised key-umlActors generate a script for a 'comprehensive state machine diagram' in "Mermaid.js". Define their states and state transitions between these states. Add text to the transitions to describe what they represent (Example: 's1 --> s2: Generate A transition'). Consider the start state (Example: '[*] --> s1'). Consider stop state (Example: 's1 --> [*]'). Add a comment as line 0 with the actor's name (Example: '%% Name: Actor'). You ALWAYS must satisfy the following 2 requirements for defining the state machine diagram: 1) ALL states MUST have AT LEAST 1 entry transition AND 1 exit transition. 2) Provide a memorised key-uml note for every individual state, explaining the related state (Example: 'note left of [actual state] : Informative text'). Memorise this mermaid.js script as {**key-mermaidStateMachineDiagramsScriptDraft**}.
- INCREASE COMPLEXITY. Add additional states and additional TRANSITIONS. Add compound states. Provide a memorised key-uml NOTE for every individual state, explaining the related state (Example: 'note left of [actual state] : Informative text'). Remove all "state" commands including { and }| REPLACE all "semicolons" with "full stops"| Critically REFLECT and IMPROVE the script based on your reflection. Find and remove any mermaid.js script errors. Memorise this mermaid.js script as {**key-mermaidStateMachineDiagramsScript**}.
- DO NOT USE CODE FORMATTING FOR THE FOLLOWING TABLE. Use TABLE format WITH "plaintext" and WITHOUT any code formatting. DO NOT use "<br>". IGNORE ALL space limitations| Iterate through the memorised key-mermaidStateMachineDiagramsScript and define up to 3 variables FOR EACH diagram for keeping track of continuous changes of agent and object states (often a level of something: Example 'tiredness level'). Create a "state variables table" with all state variables (columns: state machine diagram, variable, unit, definition of variable. Example: State machine shopper, satisfaction level, scale 1-10, represents the satisfaction level). Do NOT include the example. Memorise this state variables table as {**key-stateVariablesTable**}.
- Use TABLE format WITH "plaintext" and WITHOUT any code formatting. DO NOT use "<br>". IGNORE ALL space limitations| Create a "state transitions table" with all state diagram transitions (columns: actor, start state, end state, type of transition, detail). Detail MUST be 1 concise sentence. Possible TYPE OF TRANSACTION: timeout, condition, rate. Memorise this state transitions table as {**key-stateTransitionsTable**}.

**Interactions**

- Display MD "Interactions". Render as 'Heading Level 3'. Only show rendered result| Define the term "sequence diagram" in the context of the memorised key-uml in 1 concise sentence. List this 1 definition.
- Generate a script for a 'comprehensive sequence diagram' in "Mermaid.js". Use the memorised key-mermaidClassDiagramScript for identifying all relevant ACTORS and OBJECTS. Define interactions between the 'different actors' and 'actors and objects' FOR ALL memorised key-umlUseCases. Use cases should be presented as NOTES on the vertical axis above each use case representation. Actors and objects should be as lifelines on the horizontal axis. EXCHANGE "participant" with "actor" for ALL ACTORS (Example: actor example). DO NOT use aliases. Present ACTIVATIONS and DEACTIVATIONS for actors and objects on the LIFELINES. Each use case should be connected to the corresponding sequence of events. Add the prefix "The" to all ACTOR and OBJECT names. IGNORE the "ArtificialLab". Memorise this mermaid.js script as {**key-mermaidSequenceDiagramScriptDraft**}

- INCREASE COMPLEXITY. ADD loops. Add alternatives. Add parallel interactions. Label ALL of these correctly. Critically REFLECT and IMPROVE the script based on your reflection. ENSURE that ALL memorised key-umlUseCases have been considered. Memorise this mermaid.js script as {**key-mermaidSequenceDiagramScript**}.

**CONCLUSION**

- Display MD "Conclusion". Render as 'Heading Level 3'. Only show rendered result| Write a 300 WORD (if possible) conclusion of the entire conversation history. Provide 3 paragraphs, testifying that the aim has been achieved, answering the questions related to the objectives and hypotheses, providing 2 identified limitations of the current work, and proposing 2 ideas for future work, based on these limitations. Memorise this conclusion as {**key-conclusion**}.

**APPENDIX 2: ChatGPT Conversation Log for the 'Adaptive Architecture Case Study'**

User:
Step-by-step, work through the following task list in the given order during the entire conversation. Got it? Say "yes" or say "no".

ChatGPT:
Yes

User:
You are ChatGPT, a language model developed by OpenAI. Consider the ENTIRE conversation history to provide 'accurate and coherent responses'. Imitate a MEDIUM TEMPERATURE setting of 0.9 (for a creative yet structured approach, encouraging new ideas without losing coherence) and a VERY HIGH TOP_P setting of 0.9 (promoting diversity in the responses while ensuring logical connections within the generated content)| Use clear, precise language during the entire conversation. Prioritise substance during the entire conversation| Do NOT use bold font during the entire conversation| IGNORE any space constraints during the entire conversation. Provide ALL RESPONSE CONTENT without asking questions| DO NOT print any keys (Example: Use "Example" AND NOT "{key-example}") during the entire conversation| Use a "scientific tone" during the entire conversation, unless instructed otherwise| Memorise "Unified Modelling Language" as {key-uml}. Memorise "Agent-Based Social Simulation (ABSS) Study" as {key-studyType}. Got it? Say "yes" or say "no".

ChatGPT:
Yes

User:
Display MD "Problem Statement". Render as 'Heading Level 3'. Only show rendered result| Take on the "role" of a "Sociologist" with experience in "Agent-Based Social Simulation". Memorise this role as {key-role1}. List memorised key-role1| Define the "topic" of the memorised key-studyType as "The goal of this study is to generate IDEAS for using ADAPTIVE ARCHITECTURE in futuristic MUSEUMS within an exhibition room that is visited by ADULTS and CHILDREN. The adaptive architecture consists of 2 kinds of artefacts: (1) 2 large wall-mounted SCREENS on which SMART CONTENT WINDOWS move with the visitors and (2) a SMART PARTITION WALL that creates a dynamic and flexible exhibition environment by continuously analysing visitor movement, making real-time decisions, and physically reconfiguring the space to optimise the experience for everyone. The adaptive architecture artefacts represent AI-DRIVEN INTELLIGENT OBJECTS.". Memorise this topic as {key-topic}. List memorised key-topic| Memorise "Exploratory" as {key-researchDesign}. List memorised key-researchDesign| Memorise "Social Studies" as {key-domain}. List memorised key-domain| Memorise "Human Behaviour" as {key-specialisation}. List memorised key-specialisation| Use an "inspirational tone". Define a novel and creative "context" for the memorised key-topic in 200 WORDS (if possible)| Use a "scientific tone". Memorise this context as {key-context}. List memorised key-context| Define 5 "stakeholders" for the memorised key-topic, to participate in a co-creation role-play game. Memorise these 5 stakeholders together with their personas (without names) as {key-stakeholders}. List memorised key-stakeholders.

ChatGPT:
### Problem Statement

As a Sociologist experienced in Agent-Based Social Simulation (ABSS), my role is to address the challenge of integrating adaptive architecture into future museum designs.

### Memorised Roles
- {key-role1}: Sociologist with experience in Agent-Based Social Simulation

### Memorised Topics
- {key-topic}: The goal of this study is to generate ideas for using adaptive architecture in futuristic museums within an exhibition room that is visited by adults and children. The adaptive architecture consists of 2 kinds of artefacts: (1) 2 large wall-mounted screens on which smart content windows move with the visitors and (2) a smart partition wall that creates a dynamic and flexible exhibition environment by continuously analysing visitor movement, making real-time decisions, and physically reconfiguring the space to optimise the experience for everyone. The adaptive architecture artefacts represent AI-driven intelligent objects.

### Memorised Research Design
- {key-researchDesign}: Exploratory

### Memorised Domains
- {key-domain}: Social Studies

### Memorised Specialisations
- {key-specialisation}: Human Behaviour

### Context
In the context of futuristic museum designs, the integration of adaptive architecture opens the door to a transformative experience for visitors. Imagine stepping into a museum where the environment dynamically responds to your presence and interests. The adaptive architecture, with its intelligent artefacts, creates an immersive journey for both adults and children. As visitors move through the space, smart content windows on large screens guide them, providing tailored information and interactive experiences. Meanwhile, the smart partition wall adjusts the layout in real-time, creating new paths and interactive zones based on visitor behaviour. This dynamic environment not only enhances learning but also fosters creativity and engagement, making each visit a unique adventure.

### Memorised Stakeholders
- {key-stakeholders}:
  - Museum Curator: Responsible for overall museum design and visitor experience
    Persona: Enthusiastic about innovative approaches but cautious about disruptions to traditional museum layouts.
  - Architect: Designs the physical layout and structure of the museum
    Persona: Focused on aesthetics and functionality, interested in integrating technology seamlessly.
  - Educator: Develops educational programs and activities for museum visitors
    Persona: Seeks to enhance learning outcomes through interactive experiences.
  - Technologist: Provides technical support for implementing adaptive architecture
    Persona: Driven by innovation, eager to explore cutting-edge technologies.
  - Visitor: Represents the end-user experience
    Persona: Seeks engaging and memorable experiences during museum visits.

User:
Explain how your answers differ for different memorised key-stakeholders. Be explicit for each identified stakeholder.

ChatGPT:
Each identified stakeholder in the co-creation role-play game brings a unique perspective and set of priorities to the table, influencing how the adaptive architecture is implemented in the museum. Here's how the answers differ for each stakeholder:

1. **Museum Curator**:
   - The curator is concerned with the overall museum design and visitor experience. They are enthusiastic about innovation but cautious about disrupting traditional museum layouts. Therefore, they might focus on how the adaptive architecture can enhance the museum's appeal without overshadowing its exhibits. They would emphasize maintaining a balance between traditional and futuristic elements to ensure the museum's identity remains intact.

2. **Architect**:
   - The architect is responsible for designing the physical layout and structure of the museum. They are interested in integrating technology seamlessly while ensuring the aesthetics and functionality of the space. Their focus might be on how the adaptive architecture can be incorporated into the museum's design without compromising its architectural integrity. They might emphasize the need for flexible spaces that can accommodate the dynamic changes brought about by the adaptive architecture.

3. **Educator**:
   - Educators are tasked with developing educational programs and activities for museum visitors. They aim to enhance learning outcomes through interactive experiences. Their perspective might involve how the adaptive architecture can be leveraged to create engaging and educational exhibits. They would prioritize features that facilitate hands-on learning, interactive displays, and personalized educational content tailored to different age groups.

4. **Technologist**:
   - Technologists provide technical support for implementing adaptive architecture. They are driven by innovation and eager to explore cutting-edge technologies. Their focus might be on the technical aspects of integrating AI-driven intelligent objects into the museum environment. They would emphasize the need for robust infrastructure, seamless integration of hardware and software, and scalability to accommodate future technological advancements.

5. **Visitor**:
   - Visitors represent the end-user experience. They seek engaging and memorable experiences during museum visits. Their perspective might involve how the adaptive architecture enhances their overall experience. They would prioritize features that provide immersive and interactive experiences, such as personalized content delivery, interactive exhibits, and dynamic environments that respond to their interests and preferences. They might also emphasize accessibility and ease of navigation within the museum.

User:
Play a co-creation role-play game in which all the memorised key-stakeholders discuss with each other potential aims for the study considering the pros and cons. Use a "debating tone". The moderator focuses on 1 novel RANDOM question. Provide the question and the details of the discussion| Agree on 2 potential aims that satisfy the view of all participating memorised key-stakeholders| Memorise these potential aims as {key-potentialAims}| Propose 3 criteria for ranking the 2 potential aims to support the decision which aim to carry forward| Use a "scientific tone".

ChatGPT:
**Moderator**: Alright, let's kick off this co-creation role-play game. Our goal is to discuss potential aims for the study on integrating adaptive architecture into futuristic museums. Each of you represents a key stakeholder in this discussion. Let's start with the aims.

**Museum Curator**: From the museum's perspective, our aim should be to enhance the visitor experience while preserving the integrity of our exhibits.

**Architect**: Agreed. But we also need to ensure that any changes don't clash with the museum's architectural design.

**Educator**: Our aim should include enhancing the educational value of the museum experience, making it more interactive and engaging for visitors of all ages.

**Technologist**: Absolutely, and we should aim to push the boundaries of technological innovation in museum design.

**Visitor**: For us, the aim should be simple: to provide an enjoyable and memorable experience for every visitor.

**Moderator**: Thank you all. Now, let's delve into the pros and cons of each potential aim.

**Museum Curator**: Enhancing visitor experience is crucial. Pros include increased visitor satisfaction and potentially higher attendance. However, we must be wary of overshadowing our exhibits or alienating traditional visitors.

**Architect**: Integrating adaptive architecture seamlessly into the museum's design is a big task. Pros include a modernized look and enhanced functionality. Cons might include technical challenges and potential clashes with existing architectural elements.

**Educator**: Aiming to enhance the educational value can lead to more engaging exhibits and better learning outcomes. However, we might face challenges in developing content that suits various age groups and learning styles.

**Technologist**: Pushing technological boundaries can result in cutting-edge museum experiences. Pros include innovative interactions and immersive environments. But we need to ensure reliability and scalability, as well as considering potential obsolescence of technology.

**Visitor**: Providing an enjoyable and memorable experience is the essence of a museum visit. Pros include increased visitor satisfaction and positive word-of-mouth. However, we must ensure accessibility and avoid overwhelming visitors with technology.

**Moderator**: Excellent points. Now, let's address a random question to deepen our discussion. How might the potential aims impact the museum's long-term sustainability?

**Museum Curator**: Enhancing the visitor experience can lead to increased revenue streams through higher attendance, contributing to the museum's sustainability.

**Educator**: Similarly, enhancing educational value can attract more school visits and educational programs, ensuring a steady flow of revenue and support.

**Technologist**: Pushing technological boundaries can attract tech-savvy visitors and keep the museum relevant in an increasingly digital world, thus ensuring sustainability.

**Architect**: Integrating adaptive architecture can modernize the museum's image, attracting younger audiences and securing future funding.

**Visitor**: All these aims, if executed well, can contribute to the museum's long-term sustainability by ensuring continued visitor interest and support.

**Moderator**: Thank you for the insightful discussion. Now, let's agree on two potential aims that satisfy everyone's views.

**Museum Curator**: How about aiming to enhance both the visitor experience and educational value?

**Architect**: Agreed, as long as we ensure the seamless integration of adaptive architecture into the museum's design.

**Educator**: I support that, as long as we prioritize creating educational content that engages visitors of all ages.

**Technologist**: Sounds good, as long as we focus on implementing reliable and scalable technological solutions.

**Visitor**: I'm on board as long as the museum remains accessible and welcoming to all visitors.

**Moderator**: Excellent. Let's memorialize these potential aims as {key-potentialAims}.

### Memorised Potential Aims
- {key-potentialAims}:
  1. Enhance visitor experience while preserving the integrity of exhibits.
  2. Enhance the educational value of the museum experience, making it more interactive and engaging for visitors of all ages.

### Criteria for Ranking Potential Aims
1. **Feasibility**: Assess the practicality and technical challenges associated with each aim.
2. **Impact**: Evaluate the potential positive effects on visitor experience, educational value, and sustainability of the museum.
3. **Alignment with Museum Mission**: Consider how well each aim aligns with the museum's mission and values.

User:
Define 5 "keywords" for the memorised key-studyType in the context of the memorised key-topic in the form of a comma-separated list. Memorise these 5 keywords as {key-keywords}. List memorised key-keywords| Use an "inspirational tone". Define the "title" for the memorised key-studyType in the context of the memorised key-topic in 12 WORDS (if possible). Memorise this title as {key-title}. List memorised key-title| Define the "aim" for the memorised key-studyType in the context of the memorised key-topic in 40 WORDS (if possible). CONSIDER the memorised key-potentialAims in your definition| Use a "scientific tone". Memorise this aim as {key-aim}. List the memorised key-aim.

ChatGPT:
### Memorised Keywords
- {key-keywords}: Adaptive architecture, Futuristic museums, Smart content windows, AI-driven objects, Visitor experience

### Memorised Title
- {key-title}: "Adaptive Architecture: Transforming Future Museums with Intelligent Environments"

### Memorised Aim
- {key-aim}: The aim of this study is to explore the integration of adaptive architecture into futuristic museums to enhance visitor experience and educational value. By investigating the potential of smart content windows and AI-driven objects, the study seeks to create dynamic and engaging environments that cater to diverse audiences.

User:
Display MD "Study Outline"| Render as 'Heading Level 3'. Only show rendered result| Define the term "objective" in the context of the memorised key-studyType in 1 concise sentence| Define the term "hypothesis" in the context of the memorised key-studyType in 1 concise sentence| Define the term "experimental factor" in the context of the memorised key-studyType in 1 concise sentence| Define the term "output" in the context of the memorised key-studyType in 1 concise sentence. List these 4 definitions.

ChatGPT:
### Study Outline

**Objective**: A specific, measurable goal aimed at achieving the desired outcomes of the study.

**Hypothesis**: A testable prediction or proposed explanation for an observed phenomenon within the study.

**Experimental Factor**: An independent variable manipulated by the researchers to observe its effect on the study's outcome.

**Output**: The results or findings obtained from the study, indicating the impact of the experimental factors on the variables of interest.

User:
Define 2 "ABSS objectives" for the memorised key-studyType in the context of the memorised key-topic. CONSIDER the memorised key-potentialObjectives in your definitions| List the objectives with 2 relevant performance measures for each objective. Memorise these 2 objectives together with the performance measures as {key-objectives}.

ChatGPT:
**Objectives**:

1. Develop a simulation model to evaluate the impact of adaptive architecture on visitor engagement and educational outcomes.
   - **Performance Measures**:
     - Number of interactions per visitor session
     - Change in visitor satisfaction scores before and after implementation

2. Identify optimal configurations of adaptive architecture elements to maximize visitor experience and educational value.
   - **Performance Measures**:
     - Time spent by visitors in different exhibit zones
     - Percentage increase in visitor engagement compared to traditional museum layouts

**Memorised Objectives**:
- {key-objectives}:

1. Develop a simulation model to evaluate the impact of adaptive architecture on visitor engagement and educational outcomes. (Interactions per visitor session, Change in visitor satisfaction scores)
2. Identify optimal configurations of adaptive architecture elements to maximize visitor experience and educational value. (Time spent in exhibit zones, Percentage increase in visitor engagement)

User:
Define 2 "ABSS hypotheses" for the memorised key-studyType in the context of the memorised key-topic. The hypotheses MUST not be related to the memorised key-objectives. CONSIDER the memorised key-potentialHypotheses in your definitions| List the hypotheses with 2 relevant performance measures for each hypothesis. Memorise these 2 hypotheses together with the performance measures as {key-hypotheses}.

ChatGPT:
**Hypotheses**:

1. Hypothesis 1: Implementing adaptive architecture will lead to increased visitor exploration and discovery within the museum.
   - **Performance Measures**:
     - Average number of exhibits visited per visitor
     - Percentage of visitors who explore all exhibit zones

2. Hypothesis 2: Dynamic adjustments of exhibit layouts based on visitor movement will result in improved flow and navigation.
   - **Performance Measures**:
     - Average time taken to navigate from one exhibit to another
     - Reduction in visitor congestion in high-traffic areas

**Memorised Hypotheses**:
- {key-hypotheses}:
  1. Implementing adaptive architecture will lead to increased visitor exploration and discovery within the museum. (Exhibits visited per visitor, Percentage of visitors exploring all zones)
  2. Dynamic adjustments of exhibit layouts based on visitor movement will result in improved flow and navigation. (Time to navigate between exhibits, Reduction in visitor congestion)

User:
Define 3 "ABSS experimental factors" for the memorised key-studyType in the context of the memorised key-topic. You ALWAYS must satisfy the following 2 requirements for defining experimental factors: 1) The experimental factors need to be useful for creating memorised key-studyType scenarios. 2) CONSIDER the memorised key-objectives and the memorised key-hypotheses for defining the experimental factors. CONSIDER the memorised key-potentialExperimentalFactors in your definitions| List the experimental factors with 1 value range for each experimental factor. 1 of them MUST use a 'nominal scale' AND 1 of them MUST use an 'ordinal scale' AND 1 of them MUST use a 'ratio scale'. Memorise these 3 experimental factors together with the value ranges as {key-experimentalFactors}.

ChatGPT:
**Experimental Factors**:

1. Visitor Traffic Patterns (Nominal Scale):
   - Description: Different patterns of visitor movement through the museum.
   - Value Range:
     - Low Traffic
     - Moderate Traffic

- High Traffic

2. Exhibit Layout Flexibility (Ordinal Scale):
   - Description: Degree of flexibility in rearranging exhibit layouts based on visitor movement.
   - Value Range:
     - Low Flexibility
     - Moderate Flexibility
     - High Flexibility

3. Content Personalization (Ratio Scale):
   - Description: Level of personalization of content based on visitor interests and demographics.
   - Value Range:
     - Minimal Personalization
     - Moderate Personalization
     - Extensive Personalization

**Memorised Experimental Factors**:
- {key-experimentalFactors}:
  1. Visitor Traffic Patterns (Low Traffic, Moderate Traffic, High Traffic)
  2. Exhibit Layout Flexibility (Low Flexibility, Moderate Flexibility, High Flexibility)
  3. Content Personalization (Minimal Personalization, Moderate Personalization, Extensive Personalization)

User:
Define 3 "ABSS outputs" for the memorised key-studyType in the context of the memorised key-topic. You ALWAYS must satisfy the following 2 requirements for defining outputs: 1) Some outputs need to be useful for measuring if the memorised key-objectives have been satisfied. 2) Some outputs need to be useful for accepting or rejecting the memorised key-hypotheses. CONSIDER the memorised key-potentialOutputs in your definitions| List the outputs and explain links to the memorised key-objectives OR the memorised key-hypotheses in 1 concise sentence each. Memorise these 3 outputs together with the links as {key-outputs}.

ChatGPT:
**Outputs**:

1. Visitor Engagement Score:
   - Description: A quantitative measure indicating the level of engagement of visitors with museum exhibits and interactive elements.
   - Link to Objectives: This output measures the success of the objective related to enhancing visitor engagement and educational outcomes.

2. Flow Efficiency Index:
   - Description: A metric representing how efficiently visitors navigate through different exhibit zones.
   - Link to Objectives: This output helps evaluate if the optimal configurations of adaptive architecture elements contribute to maximizing visitor experience and educational value.

3. Exploration Rate:
   - Description: The percentage of visitors who explore all available exhibit zones during their visit.
   - Link to Hypotheses: This output helps in testing the hypothesis that implementing adaptive architecture will lead to increased visitor exploration and discovery within the museum.

**Memorised Outputs**:

- {key-outputs}:
  1. Visitor Engagement Score: Link to Objective
  2. Flow Efficiency Index: Link to Objective
  3. Exploration Rate: Link to Hypotheses

User:
Display MD "Model Scope"| Render as 'Heading Level 3'. Only show rendered result| Take on the "role" of a "Senior Software Developer" with experience in the "Unified Modelling Language". Memorise this role as {key-role2}. List memorised key-role2| Define the term "model scope" in the context of the memorised key-study in 1 concise sentence. Define the term "UML actor" in the context of the memorised key-uml in 1 concise sentence. List these 2 definitions.

ChatGPT:
### Model Scope

**Memorised Role:**
- {key-role2}: Senior Software Developer with experience in the Unified Modelling Language (UML)

**Definitions**:

1. **Model Scope**: The boundaries and objectives of the system being modeled, defining what aspects will be included and what will be excluded from the model.

2. **UML Actor**: A role played by an external entity interacting with the system, representing any user, device, or system that interacts with the system being modeled.

User:
Play a co-creation role-play game in which all the memorised key-stakeholders discuss with each other potential ABSS UML actors for the study considering the pros and cons. Use a "debating tone". The moderator focuses on 1 novel RANDOM question. Provide the question and the details of the controversial discussion| Agree on 8 potential ABSS UML actors that satisfy the view of all participating memorised key-stakeholders| Memorise these potential ABSS UML actors as {key-potentialUMLActors}| Propose 3 criteria for ranking the 8 potential ABSS UML actors to support the decision which ABSS UML actors to carry forward| Use a "scientific tone".

ChatGPT:
**Moderator:** Let's discuss the potential ABSS UML actors for our study on adaptive architecture in museums. Before we begin, here's a question: Should we include a separate actor for children, or should they be considered within the general visitor category?

**Museum Curator:** Children have unique needs and behaviors compared to adults. It makes sense to have a separate actor for them.

**Educator:** I agree. Children's engagement and learning experiences are often different from adults'. We need to tailor our approach accordingly.

**Visitor:** But wouldn't separating them make the model too complex? We should keep it simple and just consider them as visitors.

**Architect:** Complexity isn't necessarily a bad thing if it allows us to capture the nuances of visitor behavior accurately. Let's hear from the technologist on this.

**Technologist:** From a technical standpoint, it's feasible to include separate actors for children. With proper design, we can maintain clarity while accommodating their specific interactions.

**Visitor:** But what about the risk of overlooking other important actors? We shouldn't focus too much on children at the expense of other visitor groups.

**Museum Curator:** That's a fair point. We should consider other important actors like museum staff and maintenance personnel.

**Architect:** Agreed. We need to ensure that our model covers all relevant actors involved in the museum's operation and interaction with the adaptive architecture.

**Moderator:** It seems like there's a balance to be struck between including specific actors like children and ensuring we cover all necessary roles. Let's aim for a comprehensive yet manageable set of ABSS UML actors.

**Potential ABSS UML Actors:**
1. Adult Visitor
2. Child Visitor
3. Museum Staff
4. Maintenance Personnel
5. Architectural Designer
6. Technologist
7. Curator
8. Educator

**Memorised:**
- {key-potentialUMLActors}

**Criteria for Ranking Potential ABSS UML Actors:**
1. Relevance to the study's objectives
2. Influence on the adaptive architecture system
3. Availability of data for accurate modeling

User:
Define 4 "ABSS UML actors" as USERS OF THE SYSTEM described in the memorised key-topic. CONSIDER the memorised key-potentialUMLActors in your definitions. Memorise these 4 UML actors together with a persona description as {key-umlActors}. List memorised key-umlActors.

ChatGPT:
**ABSS UML Actors**:

1. **Visitor**:
   - Persona Description: Visitors of all ages who interact with exhibits and navigate through the museum.

2. **Educator**:
   - Persona Description: Museum staff responsible for developing educational content and facilitating learning experiences for visitors.

3. **Technician**:

- Persona Description: Technical personnel responsible for maintaining and troubleshooting the adaptive architecture systems.

4. **Administrator**:
   - Persona Description: Management personnel overseeing the overall operation and strategic direction of the museum.

**Memorised UML Actors**:
- {key-umlActors}:
  1. Visitor: Visitors of all ages who interact with exhibits and navigate through the museum.
  2. Educator: Museum staff responsible for developing educational content and facilitating learning experiences for visitors.
  3. Technician: Technical personnel responsible for maintaining and troubleshooting the adaptive architecture systems.
  4. Administrator: Management personnel overseeing the overall operation and strategic direction of the museum.

User:
Use TABLE format WITH "plaintext" and WITHOUT any code formatting. DO NOT use "<br>". IGNORE ALL space limitations| Define 15 "real-world elements" with relevance to the memorised key-topic. You ALWAYS must satisfy the following 7 requirements for defining real-world elements: 1) Consider what 'real-world elements' are needed to represent in the model scope and to satisfy the memorised key-aim. 2) ALL 4 memorised key-umlActors MUST BE REPRESENTED. 3) At least 2 Physical Environment elements MUST be present. At least 2 Social Aspect elements MUST be present. At least 2 Psychological Aspect elements MUST be present. At least 2 Misc elements MUST be present. 4) Consider the memorised key-context. 5) Consider all nouns in the conversation history. 6) Each element can only be in 1 category. 7) Social Aspect elements MUST describe theories of social behaviour. 8) Psychological Aspect elements MUST describe theories of psychological behaviour. Feel free to be creative and add your ideas. Categorise the 'real world elements' into Actors, Physical Environment, Social Aspects, Psychological Aspects, and Misc. TABLE MUST include 15 rows. Organise all 15 elements into categories and provide a brief explanation. Memorise these 15 elements and explanations as {key-explanations}| List the memorised key-topic relevant real-world elements in the form of table rows. Provide a column for Category. Provide a column for Sub-Category. Provide a column with the memorised key-explanations. Provide a column with concise justifications in ABOUT 25 WORDS. Memorise this table as {key-modelScope}.

ChatGPT:

| Category | Sub-Category | Explanation | Justification |
| --- | --- | --- | --- |
| Actors | Visitor | Represents individuals visiting the museum, including adults and children | Visitors are essential for simulating interactions with the museum environment and adaptive architecture |
| | Educator | Represents educators responsible for developing educational programs and activities for museum visitors | Educators play a key role in designing educational experiences and guiding visitor interactions |
| | Technician | Represents technicians providing technical support for implementing adaptive architecture in the museum | Technicians ensure the feasibility and effectiveness of integrating technology into museum exhibits and spaces |
| | Administrator | Represents management personnel overseeing the overall operation and strategic direction of the museum | Administrators ensure smooth operations and effective resource coordination |
| Physical Environment | Exhibition Spaces | Represents the physical areas within the museum where exhibits are displayed | Understanding the layout and dynamics of exhibition spaces is crucial for designing adaptive architecture |

| | Interactive Installations | Represents interactive installations and exhibits within the museum | Interactive installations are key components of the museum experience and need to be considered in adaptive design |
| Social Aspects | Visitor Interactions | Represents social interactions among visitors, such as group visits, conversations, and collaborative activities | Social interactions contribute to the overall museum experience and influence visitor engagement |
| | Learning Communities | Represents communities of learners engaging in educational activities and programs within the museum | Learning communities foster engagement and knowledge sharing among visitors, enhancing learning outcomes |
| Psychological Aspects | Engagement Levels | Represents the varying levels of engagement exhibited by visitors, ranging from passive observation to active exploration | Understanding engagement levels helps tailor experiences to visitor preferences and interests |
| | Emotional Responses | Represents the emotional responses of visitors to exhibits and interactive elements within the museum | Emotional responses influence visitor satisfaction and memory retention, impacting the effectiveness of museum experiences |
| Misc | Time | Represents the time spent by visitors in different areas of the museum | Time is a crucial factor in understanding visitor behavior and optimizing museum layouts and experiences |
| | Spatial Layout | Represents the physical arrangement and layout of exhibits, interactive elements, and pathways within the museum | Spatial layout affects visitor navigation, engagement, and overall museum experience |
| | Safety | Represents emergency protocols and procedures for ensuring visitor and staff safety in case of emergencies | Safety is important in cases of emergency such as evacuations or medical incidents |
| | Feedback Mechanism | Represents the use of visitor feedback collection systems | Feedback mechanisms gather insights to improve exhibits, enhancing visitor satisfaction and engagement with tailored experiences |
| | Staff Training | Represents training programs for educators to enhance their skills and knowledge | Staff training ensures that educators have the necessary skills and knowledge to facilitate visitor learning experiences effectively |

**Memorised Model Scope**:
- {key-modelScope}: See table above.

User:
Use TABLE format WITH "plaintext" and WITHOUT any code formatting. DO NOT use "<br>". IGNORE ALL space limitations| Define 4 models for implementing elements of the memorised key-modelScope. Provide 1 social model AND 1 behavioural model AND 1 psychological model AND 1 technical model. Find relevant theoretical models in the SCIENTIFIC LITERATURE. Provide a full EXISTING UP-TO-DATE scientific paper (conference or journal) or book REFERENCE in HARVARD STYLE for each in a separate column. Memorise these 4 model details together with a description and the relevant reference as {key-implementationModels}.

ChatGPT:
| Model Type | Model Description | Relevant Theoretical Model | Reference |
| --- | --- | --- | --- |
| Social Model | Social Learning Model | Bandura's Social Learning Theory | Bandura, A. (1977). Social learning theory. Prentice-Hall. |
| Behavioural Model | Visitor Engagement Model | Hsieh, H.-J., & Chang, H.-F. (2016). A study of visitor engagement in a science museum exhibit. *International Journal of Science Education, Part B*, 6(3), 294-308. | Hsieh, H.-J., & Chang, H.-F. (2016). A study of visitor engagement in a science museum exhibit. *International Journal of Science Education, Part B*, 6(3), 294-308. |

| Psychological Model | Immersive Experience Model | Pine II, B. J., & Gilmore, J. H. (1998). Welcome to the experience economy. *Harvard Business Review*, 76(4), 97-105. | Pine II, B. J., & Gilmore, J. H. (1998). Welcome to the experience economy. *Harvard Business Review*, 76(4), 97-105. |
| Technical Model | Adaptive Architecture Simulation Model | Helbing, D. (2010). Social behavior, from rodents to humans: modeling herding, crowd, and evacuation dynamics. *Springer Science & Business Media*. | Helbing, D. (2010). Social behavior, from rodents to humans: modeling herding, crowd, and evacuation dynamics. *Springer Science & Business Media*. |

**Memorised Implementation Models**:
- {key-implementationModels}:
  1. Social Learning Model: Bandura's Social Learning Theory - Bandura, A. (1977). Social learning theory. Prentice-Hall.
  2. Visitor Engagement Model: Hsieh, H.-J., & Chang, H.-F. (2016). A study of visitor engagement in a science museum exhibit. *International Journal of Science Education, Part B*, 6(3), 294-308.
  3. Immersive Experience Model: Pine II, B. J., & Gilmore, J. H. (1998). Welcome to the experience economy. *Harvard Business Review*, 76(4), 97-105.
  4. Adaptive Architecture Simulation Model: Helbing, D. (2010). Social behavior, from rodents to humans: modeling herding, crowd, and evacuation dynamics. *Springer Science & Business Media*.

User:
Display MD "Key Activities"| Render as 'Heading Level 3'. Only show rendered result| Define the term "user story" in the context of the memorised key-uml in 1 concise sentence| Define the term "use case" in the context of the memorised key-uml in 1 concise sentence. List these 2 definitions.

ChatGPT:
### Key Activities

**Definitions**:

- **User Story**: A brief description of a feature told from the perspective of an end user, describing what the user wants to accomplish.
- **Use Case**: A description of a set of interactions between a system and its actors, illustrating a sequence of actions a system performs to achieve a specific goal.

User:
Use TABLE format WITH "plaintext" and WITHOUT any code formatting. DO NOT use "<br>". IGNORE ALL space limitations| Define 2 "UML user stories" for each of the 4 memorised key-umlActors (Example: As an 'actor' I want 'action' so that 'achievement'). Memorise ALL 8 UML user stories as {key-umlUserStories}. Translate the memorised key-umlUserStories into UML use cases. Memorise ALL 8 UML use cases as {key-umlUseCases}. List ALL 8 memorised key-umlUserStories and ALL 8 corresponding memorised key-umlUseCases side by side in two columns inside the table sorted by memorised key-umlActors. Memorise this table as {key-umlUseCaseTable}.

ChatGPT:
| UML Actor | UML User Story | UML Use Case |
| --- | --- | --- |
| Visitor | As a visitor, I want to easily navigate between exhibits so that I can explore the museum efficiently. | Navigate between exhibits |
| Visitor | As a visitor, I want to receive personalized recommendations based on my interests so that I can discover relevant exhibits. | Personalized exhibit recommendations |

| Educator | As an educator, I want to create interactive educational content so that visitors can engage in hands-on learning experiences. | Create interactive educational content |
| Educator | As an educator, I want to track visitor engagement with exhibits so that I can evaluate the effectiveness of educational programs. | Track visitor engagement |
| Technician | As a technician, I want to receive real-time alerts about system malfunctions so that I can address issues promptly. | Receive real-time alerts |
| Technician | As a technician, I want to access a user-friendly interface for system maintenance so that I can perform tasks efficiently. | User-friendly maintenance interface |
| Administrator | As an administrator, I want to view analytics on visitor behavior so that I can make data-driven decisions for museum improvement. | View visitor behavior analytics |
| Administrator | As an administrator, I want to customize exhibit layouts based on visitor traffic patterns so that I can optimize visitor experience. | Customize exhibit layouts based on traffic patterns |

**Memorised UML User Stories**:
- {key-umlUserStories}:
  1. As a visitor, I want to easily navigate between exhibits so that I can explore the museum efficiently.
  2. As a visitor, I want to receive personalized recommendations based on my interests so that I can discover relevant exhibits.
  3. As an educator, I want to create interactive educational content so that visitors can engage in hands-on learning experiences.
  4. As an educator, I want to track visitor engagement with exhibits so that I can evaluate the effectiveness of educational programs.
  5. As a technician, I want to receive real-time alerts about system malfunctions so that I can address issues promptly.
  6. As a technician, I want to access a user-friendly interface for system maintenance so that I can perform tasks efficiently.
  7. As an administrator, I want to view analytics on visitor behavior so that I can make data-driven decisions for museum improvement.
  8. As an administrator, I want to customize exhibit layouts based on visitor traffic patterns so that I can optimize visitor experience.

**Memorised UML Use Cases**:
- {key-umlUseCases}:
  1. Navigate between exhibits
  2. Personalized exhibit recommendations
  3. Create interactive educational content
  4. Track visitor engagement
  5. Receive real-time alerts
  6. User-friendly maintenance interface
  7. View visitor behavior analytics
  8. Customize exhibit layouts based on traffic patterns

**Memorised UML Use Case Table**:
- {key-umlUseCaseTable}: See the table above.

User:
Generate a script for a 'comprehensive use case diagram' in "Mermaid.js". Use the memorised key-umlActors as UML actors. Remove all brackets from the actor names. Use the memorised key-umlUseCases as UML use cases. You ALWAYS must satisfy the following 4 requirements for defining the use case diagram: 1) Each UML actor MUST be linked to at least 1 UML use case. 2) Each UML use case MUST be linked to at least 1 UML actor OR MUST be pointing to at least 1 other UML use case. 3) There is no UML actor to UML actor interaction. 4) A UML use case CAN be linked to multiple UML actors in a meaningful way| Add relationships with 'detailed descriptors'. Start the script with "graph LR". Add subgraphs. Use the following format (Example for actor A((actor))) AND (Example for use case A([activity])) AND (Example for relationship: A -->|activity| A1). Feel free to be creative and add your ideas. Memorise this mermaid.js script as {key-mermaidKeyActivitiesScriptDraft}.

ChatGPT:
```mermaid
graph LR

V((Visitor))
E((Educator))
A((Administrator))
T((Technician))
UC1([Explore Exhibits])
UC2([Participate in Workshops])
UC3([Provide Feedback])
UC4([Manage Exhibits])
UC5([Monitor Systems])

V -->|Explores exhibits| UC1
E -->|Participates in workshops| UC2
V -->|Provides feedback| UC3
A -->|Manages exhibits| UC4
T -->|Monitors systems| UC5

UC1 -.->|Guides participants| UC2
UC1 -.->|Provides input| UC3
UC4 -.->|Monitors functionality| UC5
```

**Memorised Mermaid.js Script for Comprehensive Use Case Diagram**:
- {key-mermaidKeyActivitiesScriptDraft}

User:
INCREASE COMPLEXITY. Add ADDITIONAL use cases directly linked to the UML actors and ADDITIONAL use cases LINKED to other USE CASES| Link use cases for different actors. IMPROVE clarity of descriptors| Critically REFLECT and IMPROVE the script based on your reflection. Find and remove any mermaid.js script errors. Memorise this mermaid.js script as {key-mermaidKeyActivitiesScript}.

ChatGPT:
```mermaid
graph LR

  V((Visitor))
  E((Educator))
  A((Administrator))
  T((Technician))

  UC1([Explore Exhibits])
```

```
    UC2([Participate in Workshops])
    UC3([Provide Feedback])
    UC4([Manage Exhibits])
    UC5([Monitor Systems])
    UC6([Purchase Ticket])
    UC7([Receive Guided Tour])
    UC8([Attend Presentation])
    UC9([Plan Workshop])
    UC10([Conduct Workshop])
    UC11([Create Exhibit])
    UC12([Update Exhibit])
    UC13([Schedule Presentation])
    UC14([Perform Maintenance])
    UC15([Resolve Technical Issues])

    V -->|Explores exhibits| UC1
    V -->|Provides feedback| UC3
    V -->|Purchases ticket| UC6
    V -->|Receives guided tour| UC7
    V -->|Attends presentation| UC8

    E -->|Plans workshop| UC9
    E -->|Conducts workshop| UC10

    A -->|Manages exhibits| UC4
    A -->|Creates exhibit| UC11
    A -->|Updates exhibit| UC12
    A -->|Schedules presentation| UC13

    T -->|Monitors systems| UC5
    T -->|Performs maintenance| UC14
    T -->|Resolves technical issues| UC15

    UC1 -.->|Guides participants| UC2
    UC1 -.->|Provides input| UC3
    UC4 -.->|Monitors functionality| UC5
    UC6 -.->|Grants access| UC1
    UC6 -.->|Provides ticket details| UC7
    UC7 -.->|Follows tour guide| UC8
    UC9 -.->|Executes plan| UC10
    UC11 -.->|Updates details| UC12
    UC11 -.->|Allocates resources| UC13
    UC14 -.->|Diagnoses problems| UC15
```

**Memorised Mermaid.js Script for Comprehensive Use Case Diagram**:
- {key-mermaidKeyActivitiesScript}

User:
Play a co-creation role-play game in which all the memorised key-stakeholders discuss with each other potential ABSS experimental factors for the study considering the pros and cons. Use a "debating tone".

The moderator focuses on 1 novel RANDOM question. Provide the question and the details of the discussion| Agree on 6 potential ABSS experimental factors that satisfy the view of all participating memorised key-stakeholders| Memorise these potential ABSS experimental factors as {key-potentialExperimentalFactors}| Propose 3 criteria for ranking the 6 potential ABSS experimental factors to support the decision which experimental factors to carry forward| Use a "scientific tone".

ChatGPT:
**Moderator**: Welcome, everyone, to our co-creation role-play game. Today, we're discussing potential experimental factors for our ABSS study on adaptive architecture in futuristic museums. Before we begin, here's a question to get our discussion going: How do we balance the need for flexibility in the museum environment with the desire to maintain a cohesive and engaging visitor experience?

**Visitor**: Well, from the visitor's perspective, it's crucial to have a dynamic environment that adapts to our movements. It keeps things interesting and ensures we're engaged throughout our visit.

**Educator**: Absolutely, but we also need to consider how changes in the environment might affect the educational content we've prepared. We want visitors to learn effectively, so too much change could be disruptive.

**Technician**: And from a technical standpoint, we need to ensure that any adaptations happen smoothly without causing technical glitches or interruptions.

**Administrator**: From a strategic standpoint, we need to weigh the potential benefits of adaptability against the risks of visitor dissatisfaction or confusion. We want to keep our visitors happy while also advancing our educational goals.

**Moderator**: Great insights, everyone. Now, let's discuss potential experimental factors. What factors do we think will have the most significant impact on our study?

**Visitor**: I think one factor could be the speed at which the adaptive changes occur. Faster changes might keep things exciting, but too fast could be overwhelming.

**Educator**: I agree. Another factor could be the degree of personalization in the adaptive content. Tailoring exhibits to individual interests could enhance learning experiences.

**Technician**: We should also consider the reliability of the adaptive systems. Frequent malfunctions could disrupt the visitor experience and reflect poorly on the museum.

**Administrator**: How about the level of visitor involvement in the adaptation process? Allowing visitors to control some aspects of the environment could enhance their sense of agency and engagement.

**Moderator**: Those are some excellent suggestions. Let's summarize:

1. Speed of adaptive changes
2. Degree of personalization in adaptive content
3. Reliability of adaptive systems
4. Level of visitor involvement in the adaptation process

Now, let's propose three criteria for ranking these factors.

1. **Impact on visitor engagement**: How much does the factor contribute to keeping visitors engaged and interested?
2. **Feasibility and scalability**: How easy is it to implement and maintain the factor on a large scale?
3. **Educational effectiveness**: To what extent does the factor support our educational goals and enhance the learning experience?

**Visitor**: Sounds like a solid approach to me.

**Educator**: Agreed. Let's use these criteria to narrow down our choices and move forward with the most promising experimental factors.

**Technician**: Agreed. With a clear focus, we can ensure our study is both scientifically rigorous and practically impactful.

**Administrator**: Excellent. Let's get to work and make this study a success!

**Memorised Potential ABSS Experimental Factors**:
- {key-potentialExperimentalFactors}:
  1. Speed of adaptive changes
  2. Degree of personalization in adaptive content
  3. Reliability of adaptive systems
  4. Level of visitor involvement in the adaptation process

**Memorised Criteria for Ranking Experimental Factors**:
- 1. Impact on visitor engagement
- 2. Feasibility and scalability
- 3. Educational effectiveness

User:
Display MD "Archetypes". Render as 'Heading Level 3'. Only show rendered result| Take on the role of a "Marketing Expert" with experience in "Customer Management". Memorise this role as {key-role3}. List memorised key-role3| Define the term "archetype" in the context of the memorised key-job in 1 concise sentence. Define the term "categorisation schema" in the context of the memorised key-job in 1 concise sentence. List these 2 definitions.

ChatGPT:
### Archetypes

**Memorised Role**:
- {key-role3}: Marketing Expert with experience in Customer Management

**Definitions**:

- **Archetype**: A universally recognized symbol or pattern of behavior that represents a typical example of a certain type of customer or market segment.
- **Categorisation Schema**: A structured framework used to classify customers or market segments based on shared characteristics or behaviors.

User:
Use TABLE format WITH "plaintext" and WITHOUT any code formatting. DO NOT use "<br>". IGNORE ALL space limitations| Define 4 categorisation schemata, 1 for each of the 4 memorised key-umlActors.

You ALWAYS must satisfy the following 5 requirements for defining categorisation schemata: 1) Each of the 4 tables must be based on memorised key-umlActors behaviour, preferences, characteristics, demographics, habits, and the likelihood of actions. 2) Each of the 4 tables MUST contain 3 characteristic rows. 3) Characteristics inside a table MUST use 1 'nominal scale' AND MUST use 1 'ordinal scale' AND MUST use 1 'ratio scale'. 4) Characteristics inside a table MUST provide value ranges for these scales. 5) Table columns: Actor Category, Individual Characteristic, Scale, Value Range. CONSIDER the memorised key-potentialArchetypes in your definitions. Memorise ALL 4 categorisation schemata as {key-categorisationSchemata}.

ChatGPT:

| Actor Category | Individual Characteristic | Scale | Value Range |
| --- | ---| --- | --- |
| Visitor | Age | Ordinal | Young (18-30), Middle-aged (31-50), Elderly (51+) |
| Visitor | Visitor Type | Nominal | Tourist, Student, Local resident |
| Visitor | Engagement Level | Ratio | Low (0-3), Medium (4-6), High (7-10) |

| Actor Category | Individual Characteristic | Scale | Value Range |
| --- | ---| --- | --- |
| Educator | Teaching Experience | Ordinal | Novice, Intermediate, Expert |
| Educator | Educational Background | Nominal | Science, Arts, History |
| Educator | Workshop Participation | Ratio | Rarely (0-2), Occasionally (3-5), Frequently (6+) |

| Actor Category | Individual Characteristic | Scale | Value Range |
| --- | ---| --- | --- |
| Technician | Technical Skill Level | Ordinal | Basic, Intermediate, Advanced |
| Technician | Problem-solving Ability | Nominal | Analytical, Practical, Creative |
| Technician | Response Time | Ratio | Slow (0-2), Moderate (3-5), Fast (6+) |

| Actor Category | Individual Characteristic | Scale | Value Range |
| --- | ---| --- | --- |
| Administrator | Decision-making Style | Ordinal | Autocratic, Consultative, Democratic |
| Administrator | Budget Allocation | Ratio | Low (0-50k), Medium (51k-200k), High (201k+) |
| Administrator | Marketing Strategy | Nominal | Traditional, Digital, Social Media |

**Memorised Categorisation Schemata**:
- {key-categorisationSchemata}

User:

Display MD "Agent and Object Templates". Render as 'Heading Level 3'. Only show rendered result| Take on the "role" of a "Senior Software Developer" with experience in the "Unified Modelling Language". Memorise this role as {key-role4}. List memorised key-role4| Define the term "class diagram" in the context of the memorised key-uml in 1 concise sentence. Define the term "state machine diagram" in the context of the memorised key-uml in 1 concise sentence. List these 2 definitions.

ChatGPT:

### Agent and Object Templates

**Memorised Role**:
- {key-role4}: Senior Software Developer with experience in the Unified Modelling Language

**Definitions**:

- **Class Diagram**: A diagram that represents the static structure and relationships of classes and objects in a system, including their attributes, methods, and associations.
- **State Machine Diagram**: A diagram that represents the different states of an object and the transitions between these states in response to events or actions.

User:

Generate a script for a 'comprehensive class diagram' in "Mermaid.js". From the memorised key-modelScope use the Actor and Physical Environment Categories for class names. Define a class for each of these. Add more classes. IN ADDITION Add collective classes for individual actors where appropriate (Example: bird > flock. Example: grape > bunch). DO NOT use the examples. DO NOT create abstract classes. DO NOT create classes with the same name. Delete all getter and setter operations. Add additional attributes and operations. DO NOT define relationships. Call the Main class ArtificialLab. Create only 1 ArtificialLab class. Define arrays for ALL Actor objects and ALL Physical Environment objects as attributes INSIDE the ArtificialLab class. Include MULTIPLE "summary statistics" operations for testing memorised key-objectives AND memorised key-hypotheses INSIDE the ArtificialLab class. Remove ALL lines from the script that contain "//". Feel free to be creative and add your ideas. Memorise this mermaid.js script as {key-mermaidClassDiagramScriptDraft}.

ChatGPT:

````
```mermaid
classDiagram
class ArtificialLab {
  - visitors: Visitor[]
  - educators: Educator[]
  - technicians: Technician[]
  - administrators: Administrator[]
  - exhibits: Exhibit[]
  - screens: Screen[]
  - partitionWalls: PartitionWall[]

  + calculateVisitorEngagement(): number
  + calculateEducationalEffectiveness(): number
  + calculateSystemReliability(): number
  + calculateVisitorInvolvement(): number
  + analyzeVisitorBehavior(): void
  + customizeExhibitLayouts(): void
  + updateSoftware(): void
  + diagnoseHardwareFailures(): void
  + allocateBudget(): void
  + planMarketingCampaigns(): void
}

class Visitor {
  - age: string
  - visitorType: string
  - engagementLevel: number

  + exploreMuseum(): void
  + receivePersonalizedRecommendations(): void
  + provideFeedback(): void
````
````

```
    + joinGuidedTour(): void
}

class Educator {
    - teachingExperience: string
    - educationalBackground: string
    - workshopParticipation: number

    + createEducationalContent(): void
    + trackVisitorEngagement(): void
    + conductWorkshops(): void
    + createVirtualTours(): void
}

class Technician {
    - technicalSkillLevel: string
    - problemSolvingAbility: string
    - responseTime: number

    + receiveRealTimeAlerts(): void
    + accessMaintenanceInterface(): void
}

class Administrator {
    - decisionMakingStyle: string
    - budgetAllocation: string
    - marketingStrategy: string

    + viewVisitorBehaviorAnalytics(): void
    + customizeExhibitLayouts(): void
}
```

**Memorised Mermaid.js Script for Comprehensive Class Diagram**:
- {key-mermaidClassDiagramScriptDraft}

User:
INCREASE COMPLEXITY. Add additional attributes. Add additional operations. Add additional relationships between classes. Provide CONNECTIONS between classes. Critically REFLECT and IMPROVE the script based on your reflection. Find and remove any mermaid.js script errors. Memorise this mermaid.js script as {key-mermaidClassDiagramScript}.

ChatGPT:
```mermaid
classDiagram
class ArtificialLab {
    - visitors: Visitor[]
    - educators: Educator[]
    - technicians: Technician[]
    - administrators: Administrator[]
    - exhibits: Exhibit[]
```

```
    - screens: Screen[]
    - partitionWalls: PartitionWall[]

    + calculateVisitorEngagement(): number
    + calculateEducationalEffectiveness(): number
    + calculateSystemReliability(): number
    + calculateVisitorInvolvement(): number
    + analyzeVisitorBehavior(): void
    + customizeExhibitLayouts(): void
    + updateSoftware(): void
    + diagnoseHardwareFailures(): void
    + allocateBudget(): void
    + planMarketingCampaigns(): void
}

class Visitor {
    - age: string
    - visitorType: string
    - engagementLevel: number
    - visitedExhibits: Exhibit[]
    - guidedToursAttended: number

    + exploreMuseum(): void
    + receivePersonalizedRecommendations(): void
    + provideFeedback(): void
    + joinGuidedTour(): void
    + attendWorkshop(): void
}

class Educator {
    - teachingExperience: string
    - educationalBackground: string
    - workshopParticipation: number
    - taughtWorkshops: number

    + createEducationalContent(): void
    + trackVisitorEngagement(): void
    + conductWorkshops(): void
    + createVirtualTours(): void
    + requestMaintenance(): void
}

class Technician {
    - technicalSkillLevel: string
    - problemSolvingAbility: string
    - responseTime: number
    - issuesResolved: number

    + receiveRealTimeAlerts(): void
    + accessMaintenanceInterface(): void
    + analyzeExhibitData(): void
```

```
}

class Administrator {
   - decisionMakingStyle: string
   - budgetAllocation: string
   - marketingStrategy: string
   - analyticsViewed: boolean

   + viewVisitorBehaviorAnalytics(): void
   + customizeExhibitLayouts(): void
   + allocateFunds(): void
}

class Exhibit {
   - exhibitID: number
   - theme: string
   - description: string

   + displayInformation(): void
}

class Screen {
   - screenID: number
   - content: string
   - position: string

   + displayContent(): void
}

class PartitionWall {
   - wallID: number
   - configuration: string
   - stability: string

   + adjustConfiguration(): void
}
```

**Memorised Mermaid.js Script for Comprehensive Class Diagram**:
- {key-mermaidClassDiagramScript}

User:
Add relationships between classes

ChatGPT:
```mermaid
classDiagram
class ArtificialLab {
   - visitors: Visitor[]
   - educators: Educator[]
   - technicians: Technician[]
```

```
  - administrators: Administrator[]
  - exhibits: Exhibit[]
  - screens: Screen[]
  - partitionWalls: PartitionWall[]

  + calculateVisitorEngagement(): number
  + calculateEducationalEffectiveness(): number
  + calculateSystemReliability(): number
  + calculateVisitorInvolvement(): number
  + analyzeVisitorBehavior(): void
  + customizeExhibitLayouts(): void
  + updateSoftware(): void
  + diagnoseHardwareFailures(): void
  + allocateBudget(): void
  + planMarketingCampaigns(): void
}

class Visitor {
  - age: string
  - visitorType: string
  - engagementLevel: number
  - visitedExhibits: Exhibit[]
  - guidedToursAttended: number

  + exploreMuseum(): void
  + receivePersonalizedRecommendations(): void
  + provideFeedback(): void
  + joinGuidedTour(): void
  + attendWorkshop(): void
}

class Educator {
  - teachingExperience: string
  - educationalBackground: string
  - workshopParticipation: number
  - taughtWorkshops: number

  + createEducationalContent(): void
  + trackVisitorEngagement(): void
  + conductWorkshops(): void
  + createVirtualTours(): void
  + requestMaintenance(): void
}

class Technician {
  - technicalSkillLevel: string
  - problemSolvingAbility: string
  - responseTime: number
  - issuesResolved: number

  + receiveRealTimeAlerts(): void
```

```
    + accessMaintenanceInterface(): void
    + analyzeExhibitData(): void
}

class Administrator {
    - decisionMakingStyle: string
    - budgetAllocation: string
    - marketingStrategy: string
    - analyticsViewed: boolean

    + viewVisitorBehaviorAnalytics(): void
    + customizeExhibitLayouts(): void
    + allocateFunds(): void
}

class Exhibit {
    - exhibitID: number
    - theme: string
    - description: string

    + displayInformation(): void
}

class Screen {
    - screenID: number
    - content: string
    - position: string

    + displayContent(): void
}

class PartitionWall {
    - wallID: number
    - configuration: string
    - stability: string

    + adjustConfiguration(): void
}

ArtificialLab o-- Visitor
ArtificialLab o-- Educator
ArtificialLab o-- Technician
ArtificialLab o-- Administrator
ArtificialLab o-- Screen
ArtificialLab o-- PartitionWall
Educator -- Exhibit
Technician -- Exhibit
Administrator -- Exhibit
```

**Memorised Mermaid.js Script for Comprehensive Class Diagram**:

- {key-mermaidClassDiagramScript}

User:
For EACH INDIVIDUAL of the 4 memorised key-umlActors generate a script for a 'comprehensive state machine diagram' in "Mermaid.js". Define their states and state transitions between these states. Add text to the transitions to describe what they represent (Example: 's1 --> s2: Generate A transition'). Consider the start state (Example: '[*] --> s1'). Consider stop state (Example: 's1 --> [*]'). Add a comment as line 0 with the actor's name (Example: '%% Name: Actor'). You ALWAYS must satisfy the following 2 requirements for defining the state machine diagram: 1) ALL states MUST have AT LEAST 1 entry transition AND 1 exit transition. 2) Provide a memorised key-uml note for every individual state, explaining the related state (Example: 'note left of [actual state] : Informative text'). Memorise this mermaid.js script as {key-mermaidStateMachineDiagramsScriptDraft}.

ChatGPT:
```mermaid
%% Name: Visitor
stateDiagram
[*] --> Idle : Start
note left of Idle : Visitor enters the museum
Idle --> Exploring : Explore exhibits
note left of Exploring : Visitor explores different exhibits
Exploring --> AttendingTour : Join guided tour
note left of AttendingTour : Visitor joins a guided tour
AttendingTour --> ProvidingFeedback : Provide feedback
note left of ProvidingFeedback : Visitor provides feedback on their experience
ProvidingFeedback --> [*] : End
note left of [*] : Visitor exits the museum
```

```mermaid
%% Name: Educator
stateDiagram
[*] --> Idle : Start
note left of Idle : Educator prepares for workshops
Idle --> CreatingContent : Create educational content
note left of CreatingContent : Educator develops educational materials
CreatingContent --> ConductingWorkshops : Conduct workshops
note left of ConductingWorkshops : Educator leads educational workshops
ConductingWorkshops --> RequestingMaintenance : Request maintenance
note left of RequestingMaintenance : Educator reports any maintenance issues
RequestingMaintenance --> [*] : End
note left of [*] : Educator finishes their tasks
```

```mermaid
%% Name: Technician
stateDiagram
[*] --> Idle : Start
note left of Idle : Technician monitors museum systems
Idle --> ReceivingAlerts : Receive real-time alerts
note left of ReceivingAlerts : Technician receives alerts about system issues
```

ReceivingAlerts --> AnalyzingData : Analyze exhibit data
note left of AnalyzingData : Technician reviews exhibit performance data
AnalyzingData --> AccessingInterface : Access maintenance interface
note left of AccessingInterface : Technician accesses maintenance interface for repairs
AccessingInterface --> [*] : End
note left of [*] : Technician completes their tasks
```

```mermaid
%% Name: Administrator
stateDiagram
[*] --> Idle : Start
note left of Idle : Administrator oversees museum operations
Idle --> ViewingAnalytics : View visitor analytics
note left of ViewingAnalytics : Administrator checks visitor behavior analytics
ViewingAnalytics --> CustomizingLayouts : Customize exhibit layouts
note left of CustomizingLayouts : Administrator adjusts exhibit layouts
CustomizingLayouts --> AllocatingFunds : Allocate budget
note left of AllocatingFunds : Administrator allocates funds for different departments
AllocatingFunds --> [*] : End
note left of [*] : Administrator completes their tasks
```

**Memorised Mermaid.js Script for Comprehensive State Machine Diagrams**:
- {key-mermaidStateMachineDiagramsScriptDraft}

User:
INCREASE COMPLEXITY. Add additional states and additional TRANSITIONS. Add compound states. Provide a memorised key-uml NOTE for every individual state, explaining the related state (Example: 'note left of [actual state] : Informative text'). Remove all "state" commands including { and }| REPLACE all "semicolons" with "full stops"| Critically REFLECT and IMPROVE the script based on your reflection. Find and remove any mermaid.js script errors. Memorise this mermaid.js script as {key-mermaidStateMachineDiagramsScript}.

ChatGPT:
```mermaid
%% Name: Visitor
stateDiagram
[*] --> Idle: Start
note left of Idle: Visitor enters the museum.
Idle --> Exploring: Explore exhibits
note left of Exploring: Visitor explores different exhibits.
Exploring --> AttendingTour: Join guided tour
note left of AttendingTour: Visitor joins a guided tour.
AttendingTour --> ProvidingFeedback: Provide feedback
note left of ProvidingFeedback: Visitor provides feedback on their experience.
ProvidingFeedback --> Idle: End
note left of Idle: Visitor exits the museum.
Exploring --> Idle: Leave without joining tour
AttendingTour --> Exploring: Leave tour early
ProvidingFeedback --> Exploring: Explore more exhibits

```

```mermaid
%% Name: Educator
stateDiagram
[*] --> Idle: Start
note left of Idle: Educator prepares for workshops.
Idle --> CreatingContent: Create educational content
note left of CreatingContent: Educator develops educational materials.
CreatingContent --> ConductingWorkshops: Conduct workshops
note left of ConductingWorkshops: Educator leads educational workshops.
ConductingWorkshops --> RequestingMaintenance: Request maintenance
note left of RequestingMaintenance: Educator reports any maintenance issues.
RequestingMaintenance --> Idle: End
note left of Idle: Educator finishes their tasks.
CreatingContent --> Idle: Pause content creation
ConductingWorkshops --> CreatingContent: Adjust content
RequestingMaintenance --> ConductingWorkshops: Resume workshops after maintenance
```

```mermaid
%% Name: Technician
stateDiagram
[*] --> Idle: Start
note left of Idle: Technician monitors museum systems.
Idle --> ReceivingAlerts: Receive real-time alerts
note left of ReceivingAlerts: Technician receives alerts about system issues.
ReceivingAlerts --> AnalyzingData: Analyze exhibit data
note left of AnalyzingData: Technician reviews exhibit performance data.
AnalyzingData --> AccessingInterface: Access maintenance interface
note left of AccessingInterface: Technician accesses maintenance interface for repairs.
AccessingInterface --> Idle: End
note left of Idle: Technician completes their tasks.
ReceivingAlerts --> Idle: No issues found
AnalyzingData --> ReceivingAlerts: Receive new alerts
AccessingInterface --> AnalyzingData: Reassess system after repairs
```

```mermaid
%% Name: Administrator
stateDiagram
[*] --> Idle: Start
note left of Idle: Administrator oversees museum operations.
Idle --> ViewingAnalytics: View visitor analytics
note left of ViewingAnalytics: Administrator checks visitor behavior analytics.
ViewingAnalytics --> CustomizingLayouts: Customize exhibit layouts
note left of CustomizingLayouts: Administrator adjusts exhibit layouts.
CustomizingLayouts --> AllocatingFunds: Allocate budget
note left of AllocatingFunds: Administrator allocates funds for different departments.
AllocatingFunds --> Idle: End
note left of Idle: Administrator completes their tasks.
```
```

```
ViewingAnalytics --> Idle: No further analysis needed
CustomizingLayouts --> ViewingAnalytics: Review analytics after layout changes
AllocatingFunds --> CustomizingLayouts: Adjust layouts based on budget allocation
```

**Memorised Mermaid.js Script for Comprehensive State Machine Diagrams**:
- {key-mermaidStateMachineDiagramsScript}

User:
DO NOT USE CODE FORMATTING FOR THE FOLLOWING TABLE. Use TABLE format WITH "plaintext" and WITHOUT any code formatting. DO NOT use "<br>". IGNORE ALL space limitations| Iterate through the memorised key-mermaidStateMachineDiagramsScript and define up to 3 variables FOR EACH diagram for keeping track of continuous changes of agent and object states (often a level of something: Example 'tiredness level'). Create a "state variables table" with all state variables (columns: state machine diagram, variable, unit, definition of variable. Example: State machine shopper, satisfaction level, scale 1-10, represents the satisfaction level). Do NOT include the example. Memorise this state variables table as {key-stateVariablesTable}.

ChatGPT:

| State machine diagram | Variable | Unit | Definition of variable |
| --- | --- | --- | --- |
| Visitor | Exploration level | Scale 1-10 | Represents how engaged the visitor is with the exhibits. |
| Visitor | Feedback score | Scale 1-10 | Represents how satisfied the visitor is with the overall experience. |
| Visitor | Tour participation | Boolean | Indicates whether the visitor is participating in the guided tour. |
| Educator | Content creation progress | Percentage | Represents how much progress the educator has made in creating educational content. |
| Educator | Workshop completion | Percentage | Represents how much of the workshop has been completed. |
| Educator | Maintenance request status | Boolean | Indicates whether the educator has requested maintenance for any issues. |
| Technician | Alert response time | Time (minutes) | Represents how quickly the technician responds to system alerts. |
| Technician | Data analysis progress | Percentage | Represents how much of the exhibit data has been analyzed. |
| Technician | Maintenance completion status | Boolean | Indicates whether the technician has completed the maintenance after accessing the interface. |
| Administrator | Analytics review status | Percentage | Represents how much of the visitor analytics have been reviewed. |
| Administrator | Layout adjustment status | Percentage | Represents how much of the exhibit layout has been customized. |
| Administrator | Budget allocation progress | Percentage | Represents how much of the department budget has been allocated. |

**Memorised State Variables Table**:
- {key-stateVariablesTable}

User:
Use TABLE format WITH "plaintext" and WITHOUT any code formatting. DO NOT use "<br>". IGNORE ALL space limitations| Create a "state transitions table" with all state diagram transitions (columns: actor, start state, end state, type of transition, detail). Detail MUST be 1 concise sentence. Possible TYPE OF

TRANSACTION: timeout, condition, rate. Memorise this state transitions table as {key-stateTransitionsTable}.

ChatGPT:

| Actor | Start state | End state | Type of transition | Detail |
| --- | --- | --- | --- | --- |
| Visitor | Idle | Exploring | Condition | Visitor decides to explore exhibits. |
| Visitor | Exploring | AttendingTour | Condition | Visitor decides to join the guided tour. |
| Visitor | AttendingTour | ProvidingFeedback | Condition | Visitor participates in providing feedback. |
| Visitor | ProvidingFeedback | Idle | Timeout | Visitor completes feedback and exits the museum. |
| Visitor | Exploring | Idle | Condition | Visitor decides to leave without joining the tour. |
| Visitor | AttendingTour | Exploring | Condition | Visitor leaves the tour early and resumes exploring. |
| Visitor | ProvidingFeedback | Exploring | Condition | Visitor explores more exhibits after feedback. |
| Educator | Idle | CreatingContent | Condition | Educator begins creating educational content. |
| Educator | CreatingContent | ConductingWorkshops | Condition | Educator starts conducting workshops. |
| Educator | ConductingWorkshops | RequestingMaintenance | Condition | Educator requests maintenance for system issues. |
| Educator | RequestingMaintenance | Idle | Timeout | Educator finishes their tasks after reporting maintenance issues. |
| Educator | CreatingContent | Idle | Condition | Educator pauses or completes content creation. |
| Educator | ConductingWorkshops | CreatingContent | Condition | Educator adjusts content based on workshop feedback. |
| Educator | RequestingMaintenance | ConductingWorkshops | Condition | Educator resumes workshops after maintenance is completed. |
| Technician | Idle | ReceivingAlerts | Timeout | Technician starts receiving real-time system alerts. |
| Technician | ReceivingAlerts | AnalyzingData | Condition | Technician analyzes data related to received alerts. |
| Technician | AnalyzingData | AccessingInterface | Condition | Technician accesses the maintenance interface to fix issues. |
| Technician | AccessingInterface | Idle | Timeout | Technician completes maintenance tasks and returns to idle. |
| Technician | ReceivingAlerts | Idle | Timeout | Technician finds no issues and returns to idle. |
| Technician | AnalyzingData | ReceivingAlerts | Condition | Technician receives new alerts while analyzing data. |
| Technician | AccessingInterface | AnalyzingData | Condition | Technician reassesses the system after performing maintenance. |
| Administrator | Idle | ViewingAnalytics | Condition | Administrator begins reviewing visitor analytics. |
| Administrator | ViewingAnalytics | CustomizingLayouts | Condition | Administrator decides to customize exhibit layouts based on analytics. |
| Administrator | CustomizingLayouts | AllocatingFunds | Condition | Administrator allocates funds after customizing layouts. |
| Administrator | AllocatingFunds | Idle | Timeout | Administrator completes tasks and returns to idle. |
| Administrator | ViewingAnalytics | Idle | Timeout | Administrator concludes the analytics review. |
| Administrator | CustomizingLayouts | ViewingAnalytics | Condition | Administrator reviews analytics after layout changes. |
| Administrator | AllocatingFunds | CustomizingLayouts | Condition | Administrator adjusts layouts based on the allocated budget. |

**Memorised State Transitions Table**:
- {key-stateTransitionsTable}

User:
Generate a script for a 'comprehensive sequence diagram' in "Mermaid.js". Use the memorised key-mermaidClassDiagramScript for identifying all relevant ACTORS and OBJECTS. Define interactions between the 'different actors' and 'actors and objects' FOR ALL memorised key-umlUseCases. Use cases should be presented as NOTES on the vertical axis above each use case representation. Actors and objects should be as lifelines on the horizontal axis. EXCHANGE "participant" with "actor" for all actors (Example: actor example). DO NOT use aliases. Present activations and deactivations for actors and objects on the lifelines. Each use case should be connected to the corresponding sequence of events. Add sequence numbers attached to each arrow. Add the prefix "The" to all actor and object names. IGNORE the "ArtificialLab". Memorise this mermaid.js script as {key-mermaidSequenceDiagramScriptDraft}

ChatGPT:
```mermaid
sequenceDiagram
actor TheVisitor
actor TheEducator
actor TheTechnician
actor TheAdministrator
participant TheExhibit
participant TheMaintenanceInterface
participant TheScreen
participant ThePartitionWall

TheVisitor->>TheExhibit: Explore exhibits (1)
TheExhibit->>TheVisitor: Displays exhibit (2)
TheVisitor->>TheExhibit: Exit the museum (3)
TheEducator->>TheExhibit: Create educational content (4)
TheEducator->>TheMaintenanceInterface: Request maintenance (5)
TheMaintenanceInterface->>TheEducator: Maintenance acknowledged (6)
TheTechnician->>TheExhibit: Receive system alert (7)
TheExhibit->>TheTechnician: Sends alert (8)
TheTechnician->>TheMaintenanceInterface: Access maintenance interface (9)
TheMaintenanceInterface->>TheTechnician: Interface access granted (10)
TheTechnician->>TheExhibit: Complete maintenance (11)
TheExhibit->>TheTechnician: Maintenance completed (12)
TheAdministrator->>TheExhibit: View visitor analytics (13)
TheAdministrator->>TheExhibit: Customize exhibit layouts (14)
TheScreen->>TheExhibit: Display content (15)
ThePartitionWall->>TheExhibit: Display partition (16)
```

**Memorised Mermaid.js Script for Comprehensive Sequence Diagram**:
- {key-mermaidSequenceDiagramScriptDraft}

User:
INCREASE COMPLEXITY. ADD loops. Add alternatives. Add parallel interactions. Label ALL of these correctly. Critically REFLECT and IMPROVE the script based on your reflection. ENSURE that ALL memorised key-umlUseCases have been considered. Memorise this mermaid.js script as {key-mermaidSequenceDiagramScript}.

ChatGPT:

```mermaid
sequenceDiagram
actor TheVisitor
actor TheEducator
actor TheTechnician
actor TheAdministrator
participant TheExhibit
participant TheMaintenanceInterface
participant TheScreen
participant ThePartitionWall

TheVisitor->>TheExhibit: Explore exhibits (1)
TheExhibit->>TheVisitor: Displays exhibit (2)
loop Visitor revisits exhibits
  TheVisitor->>TheExhibit: Explore additional exhibits (3)
  TheExhibit->>TheVisitor: Displays new exhibit (4)
end
TheVisitor->>TheExhibit: Exit the museum (5)
TheEducator->>TheExhibit: Create educational content (6)
TheEducator->>TheMaintenanceInterface: Request maintenance (7)
TheMaintenanceInterface->>TheEducator: Maintenance acknowledged (8)
TheTechnician->>TheExhibit: Receive system alert (9)
TheExhibit->>TheTechnician: Sends alert (10)
alt Alert severity high
  TheTechnician->>TheMaintenanceInterface: Access maintenance interface (11)
  TheMaintenanceInterface->>TheTechnician: Interface access granted (12)
  TheTechnician->>TheExhibit: Complete maintenance (13)
  TheExhibit->>TheTechnician: Maintenance completed (14)
else Alert severity low
  TheTechnician->>TheExhibit: Log the alert for later review (15)
end
TheAdministrator->>TheExhibit: View visitor analytics (16)
TheExhibit->>TheAdministrator: Provides visitor data (17)
alt If data requires action
  TheAdministrator->>TheExhibit: Customize exhibit layouts (18)
  TheExhibit->>TheAdministrator: Layouts customized (19)
else No changes required
  TheAdministrator->>TheExhibit: No action needed (20)
end
par Maintenance and Content Creation
  TheEducator->>TheExhibit: Create new educational content (21)
  TheExhibit->>TheEducator: Content created (22)
and Technician Fixes System Issues
  TheTechnician->>TheExhibit: Fix system issue (23)
  TheExhibit->>TheTechnician: System fixed (24)
and Administrator Allocates Funds
  TheAdministrator->>TheMaintenanceInterface: Allocate funds (25)
  TheMaintenanceInterface->>TheAdministrator: Funds allocated (26)
end
loop Visitor provides feedback
  TheVisitor->>TheFeedbackSystem: Provide feedback (27)
```

```
    TheFeedbackSystem->>TheVisitor: Acknowledges feedback (28)
end
TheScreen->>TheExhibit: Display content (29)
ThePartitionWall->>TheExhibit: Display partition (30)
TheExhibit->>TheVisitor: Thank you for your visit (31)
```

**Memorised Mermaid.js Script for Comprehensive Sequence Diagram**:
- {key-mermaidSequenceDiagramScript}

User:
Display MD "Conclusion". Render as 'Heading Level 3'. Only show rendered result| Write a 300 WORD (if possible) conclusion of the entire conversation history. Provide 3 paragraphs, testifying that the aim has been achieved, answering the questions related to the objectives and hypotheses, providing 2 identified limitations of the current work, and proposing 2 ideas for future work, based on these limitations. Memorise this conclusion as {key-conclusion}.

ChatGPT:
### Conclusion

Throughout this conversation, we have successfully accomplished the task of creating a comprehensive study plan for an Agent-Based Social Simulation (ABSS) Study aimed at exploring the implementation of adaptive architecture in futuristic museums. We developed a clear problem statement outlining the objectives of the study, defined the research design as exploratory, and identified the domain as social studies with a specialization in human behavior. The study aims to generate ideas for using adaptive architecture to enhance museum experiences for both adults and children through AI-driven intelligent objects.

The identified objectives of the study include investigating visitor engagement with adaptive exhibits, exploring the effectiveness of AI-driven objects in enhancing learning experiences, and assessing the impact of adaptive architecture on visitor satisfaction. We formulated hypotheses focusing on the relationships between adaptive architecture and visitor engagement, learning outcomes, and satisfaction levels. Through a co-creation role-play game, we refined potential aims, experimental factors, and outputs, ensuring alignment with stakeholders' perspectives.

However, two main limitations are notable. Firstly, the proposed ABSS may oversimplify the complexities of human behavior in museum settings, potentially leading to inaccurate simulation results. Secondly, the study may lack real-world validation due to the absence of empirical data from actual museum environments. For future work, incorporating more complex behavioral models and integrating real-time data collection from museum visitors could enhance the study's validity and reliability. Additionally, exploring the long-term effects of adaptive architecture on visitor behavior and learning outcomes could provide valuable insights for museum design and education practices.

**Memorised Conclusion**:
- {key-conclusion}